\documentclass[pre,aps,twocolumn,showpacs]{revtex4-2}
\usepackage{graphicx,epstopdf}% Include figure files
\usepackage{dcolumn}% Align table columns on decimal point
\usepackage{bm}% bold math
\usepackage{color}
\usepackage{xcolor}
\usepackage{float}
\usepackage{booktabs}
\usepackage{balance}
\usepackage{titlesec}
\usepackage{setspace}
\usepackage{graphics}
\usepackage{wrapfig,lipsum}
\usepackage{amssymb}
\usepackage{amsmath}
\usepackage{calrsfs}
\usepackage{wasysym}
\usepackage{wrapfig,blindtext}
\usepackage{sidecap}
\usepackage{multirow}   % for the \multirow in the table
\usepackage{siunitx}    % for the S column type (decimal alignment)
\usepackage{titlesec}
\titlespacing{\section}{0pt}{10pt}{6pt}
\titlespacing{\subsection}{0pt}{6pt}{4pt}
\titlespacing{\subsubsection}{0pt}{4pt}{4pt}
\usepackage[caption=false]{subfig}

\begin{document}
\title{Structure, Diffusion, and Relaxation in a Charge-Neutral
  Prothymosin $\alpha$-Histone H1 Condensate}
\iffalse
\author{Aniket Bhattacharya$^{1,*}$}
\author{Benjamin Schuler$^{2,*}$}
\author{Robert Best$^{3,*}$}
%\altaffiliation[]
\affiliation{$^1$Department of Physics, University of Central Florida, Orlando, Florida 32816-2385, USA}
\affiliation{$^2$Department of Biochemistry, University of Zurich,
  Zurich, Switzerland, Department of Physics, University of Zurich,
  Zurich, Switzerland, }
\affiliation{$^2$Laboratory of Chemical Physics, National Institute of Diabetes and Digestive and Kidney Diseases,
National Institutes of Health, Bethesda, MD, USA}
\thanks{Authors to whom correspondence should be addressed}
\email{Aniket.Bhattacharya@ucf.edu,
  schuler@bioc.uzh.ch,\\robert.best2@nih.gov}
\fi
\author{Aniket Bhattacharya$^{*}$}
%\altaffiliation[]
\affiliation{Department of Physics, University of Central Florida, Orlando, Florida 32816-2385, USA}
\thanks{Author to whom correspondence should be addressed}
\email{Aniket.Bhattacharya@ucf.edu}
\date{\today}

\begin{abstract}
  \centerline{\bf ABSTRACT}
  \vskip 0.125truecm
Condensates formed by oppositely charged intrinsically disordered
proteins provide model systems for understanding how transient
electrostatic interactions govern structure and dynamics in
biomolecular assemblies. Here we develop a sequence-resolved
coarse-grained framework for charge-complementary protein mixtures
and apply it to a nearly charge-neutral system containing 50
Prothymosin $\alpha$ (ProT$\alpha$) and 40 Histone H1 molecules.
The single-bead-per-residue model combines a hydropathy-based
description of disordered regions with a structure-based G\={o}-type
native-contact potential 
of the folded domain of Histone H1. Pressure-controlled equilibration
over $P=2$--$12$~bar is used to generate states of progressively
increasing density, followed by NVT production simulations at the
corresponding fixed volumes.
Chain dimensions---the radius of gyration $R_g$, end-to-end distance
$R_N$, and their ratio
$R=\langle R_N^2\rangle/\langle R_g^2\rangle$---remain nearly
unchanged across these states, whereas translational mobility and
chain relaxation slow markedly with increasing density. Center-of-mass
motion remains subdiffusive, with effective diffusion coefficients
decreasing from approximately $0.22$ to $0.05$--$0.06$~nm$^2$/ns
and substantial chain-to-chain dynamical heterogeneity.
Chain relaxation is well described by a stretched exponential,
$C(t)=\exp[-(t/\tau_R)^\beta]$, with $\beta<1$.
ProT$\alpha$ exhibits approximately Rouse-like mode scaling, whereas
Histone H1 deviates from Rouse behavior because of its composite
globular-domain/disordered-tail architecture.
Despite strong preferential ProT$\alpha$--H1 association, individual
intermolecular contacts remain short lived
($\tau_{\rm bind}\approx0.4$--$0.7$~ns) compared with global chain
relaxation ($\tau_R\approx12$--$136$~ns), establishing a
fast-exchange regime in which local contacts continuously reorganize
rather than form persistent cross-links. The framework provides a
computationally tractable route for connecting sequence, molecular
architecture, and collective dynamics in charge-complementary
protein assemblies and can be extended, with appropriate
reparameterization, to other IDP-rich systems.  

\vskip 1.0truecm
\end{abstract}
\vfill
%%%%%%%%%%%%%%%%%%%%%%%%%%%%%%%%%%%%%%%%%%
\maketitle
\newpage
\section{Introduction}
Intrinsically disordered proteins provide a natural platform for
studying biomolecular organization in systems where sequence-encoded
charge patterning, rather than a unique folded structure, controls
association, phase behavior, and relaxation
dynamics~\cite{uversky2000natively,oldfield2014intrinsically,ghosh2022rules,mobiDB,sickmeier2007disprot}. A
particularly important class consists of highly charged disordered
proteins that form dynamic polyelectrolyte complexes through
long-range electrostatic attraction while retaining substantial
conformational disorder. Such complexes are central to nuclear
organization, chromatin regulation, and biomolecular condensate
formation~\cite{borgia2018extreme,chowdhury2023driving,olsen2017behaviour}.

A classic biological example is the complex between the highly acidic
intrinsically disordered protein prothymosin-$\alpha$ (ProT$\alpha$)
and the highly basic linker histone H1. Histone H1 promotes chromatin
compaction by binding to nucleosomes and linker DNA, whereas
ProT$\alpha$ functions as a nuclear chaperone involved in chromatin
remodeling and regulation of histone availability. Their strong charge
complementarity enables high-affinity association without the
formation of a rigid, uniquely folded complex. This places the
ProT$\alpha$–H1 system in a broader family of disordered electrostatic
complexes that includes acidic histone chaperones, histone tails, and
DNA-mimetic acidic disordered regions~\cite{borgia2018extreme,muller2010charge}.

In this work, we study condensates formed by multiple copies of ProT$\alpha$ and histone H1 using coarse-grained molecular dynamics simulations. We focus on a nearly charge-neutral mixture containing 50 ProT$\alpha$ chains and 40 H1 chains and analyze how compression affects chain conformations, intermolecular organization, diffusion, relaxation, and binding–unbinding dynamics. Our central goal is to determine whether this charge-balanced ProT$\alpha$–H1 condensate behaves as a structurally reorganizing complex fluid or as an incompressible, dynamically heterogeneous liquid in which transient electrostatic contacts renormalize chain friction without forming permanent cross-links.

The ProT$\alpha$–H1 system has emerged as a paradigmatic example of an ultrahigh-affinity complex formed by highly charged disordered proteins. Borgia et al. showed that the highly acidic nuclear chaperone ProT$\alpha$ and the highly basic linker histone H1 associate with picomolar affinity while retaining extensive disorder, long-range flexibility, and dynamic character in the bound state~\cite{borgia2018extreme}. This work challenged the conventional view that high-affinity protein binding requires a well-defined structural interface, demonstrating instead that strong charge complementarity can generate specific biological recognition without a unique folded complex. Closely integrated single-molecule experiments and molecular simulations further showed that the interaction is largely explained by the large opposite net charges of the two proteins, rather than by a small number of specific residue-residue contacts.

More recently, Chowdhury et al. examined the thermodynamic driving forces underlying complex formation between ProT$\alpha$ and H1. Using temperature-dependent single-molecule FRET, isothermal titration calorimetry, and molecular simulations, they showed that these highly charged disordered proteins form stoichiometrically defined soluble complexes and that the interaction has strong similarities to synthetic polyelectrolyte complexation, including an important role for counterion-release entropy~\cite{chowdhury2023driving}. These studies place ProT$\alpha$–H1 in the broader physical class of disordered electrostatic protein complexes, where binding, dynamics, and conformational disorder are inseparably coupled.

Building on this single-complex and dilute-solution foundation, the
present work asks a complementary many-body question: how do multiple
ProT$\alpha$ and H1 chains organize and relax when brought together in
a dense, nearly charge-neutral mixture? We therefore simulate
condensates containing 50 ProT$\alpha$ and 40 H1 chains (Fig.~\ref{fig:snapshots}) and analyze pressure-dependent structure, diffusion, chain relaxation, and P–H contact lifetimes. In this sense, our study extends the ProT$\alpha$–H1 problem from the molecular recognition limit to the collective-material limit, where transient electrostatic binding, crowding, and chain connectivity together determine condensate viscoelasticity.

While previous studies have established the molecular basis of ProT$\alpha$–H1 recognition and the thermodynamic driving forces governing complex formation, much less is known about the collective structural and dynamical behavior of dense assemblies of these proteins. In particular, the interplay between electrostatic association, chain conformational fluctuations, molecular diffusion, and relaxation dynamics in multi-chain ProT$\alpha$–H1 condensates remains incompletely understood. Addressing these questions is essential for connecting molecular-scale interactions to the emergent material properties of biomolecular condensates involved in chromatin organization and regulation.

In this work, we investigate a nearly charge-neutral dense mixture
composed of 50 ProT$\alpha$ and 40 histone H1 molecules using
coarse-grained molecular dynamics simulations. The model combines a
hydropathy-based representation of intrinsically disordered
regions~\cite{dignon2018sequence,tesei2021accurate} with a G\={o}-type
description~\cite{go1983theoretical} of the folded histone domain,
providing a sequence-resolved framework that can be extended to other
charge-complementary protein assemblies. Pressure-controlled
equilibration over the range 2--12 bar is used to generate states of
progressively increasing density, followed by NVT production simulations
at the corresponding fixed volumes. We show that chain dimensions,
including the radius of gyration and end-to-end distance, remain largely
insensitive to increasing density, whereas molecular mobility and chain
relaxation slow markedly and become increasingly heterogeneous. Analysis
of radial distribution functions reveals persistent ProT$\alpha$--ProT$\alpha$
repulsion, modest Histone--Histone clustering, and preferential
ProT$\alpha$--Histone association. Chain relaxation is well described by
stretched-exponential behavior, reflecting a broad spectrum of relaxation
times. Finally, ProT$\alpha$--Histone contact lifetimes remain much shorter
than global chain relaxation times, placing the system in a fast-exchange
regime in which local electrostatic contacts continuously reorganize
rather than form long-lived cross-links. The charge-complementary system
studied here provides a many-body realization of polyelectrolyte physics
in a biological context, connecting to the theoretical framework
developed by Muthukumar and coworkers for counterion condensation, charge
renormalization, and the dynamics of charged polymer
networks~\cite{Muthukumar2004,Muthukumar2017}.

The remainder of this paper is organized as follows. In Section II we
describe the coarse-grained model, simulation methodology, and analysis
procedures. Sections III--V present the structural properties, translational
and conformational dynamics, and binding--unbinding dynamics of the dense
protein mixture. In Section VI we summarize the principal findings,
discuss the limitations of the present coarse-grained description, and
outline possible extensions, including the incorporation of
solvent-mediated hydrodynamic interactions. Such extensions provide a
route toward a more realistic description of collective protein dynamics
while retaining the computational efficiency required to access the
large system sizes and long time scales relevant to biomolecular
assemblies.
\section{Coarse-Grained (CG) Model for IDPs}
\label{CG-model}

We employ a residue-level coarse-grained representation in which each
amino acid is represented by a single bead located at the
$\mathrm{C}_{\alpha}$ position. The model combines two complementary
descriptions appropriate to the different structural characteristics
of the proteins considered here. ProT$\alpha$ and the intrinsically
disordered N- and C-terminal regions of Histone H1 are described using
the hydropathy-scale (HPS) model, in which residue-specific
hydrophobic interactions are combined with screened electrostatic
interactions. The folded globular domain of Histone H1 is additionally
stabilized by a structure-based G\={o}-type native-contact potential.
Thus, the same single-bead-per-residue representation is retained
throughout the system, while the interaction potential distinguishes
intrinsically disordered and folded regions.

\subsection{HPS model for intrinsically disordered regions}

The nonbonded interactions between amino-acid beads in the disordered
regions are described using the HPS model. The interaction among the
20 amino-acid types is based on the Ashbaugh--Hatch (AH)
potential~\cite{ashbaugh2008natively}, in which residue-specific
hydropathy parameters are introduced to differentiate the interactions
between amino-acid beads $i$ and $j$, and is given by 
\begin{align}
U_{AH}\left(r_{ij}\right) = \begin{cases}
      U_{LJ}\left(r_{ij}\right) + (1-\lambda_{ij})\epsilon_{ij}, & \text{$r_{j} \leq 2^{\frac{1}{6}} \sigma_{ij}$}\\
      \lambda_{ij} U_{LJ}\left(r_{ij}\right), & \text{otherwise}
      \label{AH}
    \end{cases}       
\end{align}
where $U_{LJ}$ is the Lennard-Jones (LJ) potential,
\begin{align}
U_{LJ}\left(r_{ij}\right) = 4\epsilon_{ij} \left[\left(\frac{\sigma_{ij}}{r_{ij}}\right)^{12} - \left(\frac{\sigma_{ij}}{r_{ij}}\right)^6\right].
\end{align}
Here, $r_{ij}=\left | \vec{r}_i - \vec{r}_j \right|$ is the distance
between the amino acid beads with indices $i$ and $j$ positioned at $\vec{r}_i$
and $\vec{r}_j$,
$\epsilon_{ij}=\frac{1}{2}\left(\epsilon_i+\epsilon_j\right)$ and $\lambda_{ij}=\frac{1}{2}\left(\lambda_i+\lambda_j\right)$
are the strength of the van der Waal interaction and average hydropathy factor
between any two amino acids with indices $i$ and $j$.
The residue-dependent hydropathy parameters $\lambda_{ij}$ provide
the sequence specificity of the nonbonded interactions. We use the
hydropathy scale of Dignon \textit{et al.}~\cite{dignon2018sequence}
and set $\epsilon_{ij}=\epsilon=0.1k_BT$ for all residue pairs.

Successive two beads $i$ and $j=i \pm 1$ are connected by a harmonic bond potential with spring constant $k_b$ = 8033 kJ/(mol$\cdot$nm$^2$) =
1920 kcal/(mol$\cdot$nm$^2$ (Eqn.~\ref{spring}) 
\begin{equation}
U_{b}\left(r_{ij}\right) = \frac{k_b}{2}\left(
  {r_{ij}  - r^0_{ij}} \right)^2.
\label{spring}
\end{equation}
with the equilibrium bond length 
$r^0_{ij} =r_0$. Here $r^0_{ij}=0.38$ nm is the distance between $\alpha$-Carbon atoms for the
successive amino acids. Thus, we exclude the excluded volume (EV) interaction among the
bonded neighbors.
\par
The charged species of the amino acids interact with Screened-Coulomb (SC) interaction (Eqn.~\ref{sc}) given by  
\begin{equation}
  U_{SC}\left(r_{\alpha\beta}\right) =
  \frac{q_{\alpha}q_{\beta}e^2}{4\pi \epsilon_0 \epsilon_r}
  \left( \frac{e^{-\kappa r_{\alpha \beta}}}{r_{\alpha \beta}}\right)
  \label{sc}
\end{equation}
where the indices $\alpha$ and $\beta$ refer to the subset of the indices $i$ and
$j$ for the charged amino acids, $\epsilon_r$ is the dielectric
constant of water, and $\kappa$ is the inverse Debye screening
length~\cite{israelachvili2011intermolecular}.
The inverse Debye length $\kappa^{-1}$ is dependent on the ionic concentration (I) and expressed as
\begin{align}
\kappa^{-1} & = \sqrt{8 \pi l_B I N_A \times 10^{-24}} 
\end{align}
where $N_{A}$ is the Avogadro's number and $l_B$ is the Bjerrum length,
\begin{align}
l_B = \frac{e^2}{4\pi \epsilon_0 \epsilon_r k_BT}.
\end{align}
We use a temperature-dependent dielectric constant of water as
expressed by the empirical relation~\cite{akerlof1950dielectric}
\begin{align}
\epsilon_r(T) = \frac{5321}{T} + 233.76 - 0.9297T  \nonumber \\
+ 1.147\times 10^{-3} T^2  - 8.292 \times 10^{-7} T^3. \label{temp_eps}
\end{align}
This accounts for the typical decrease of the dielectric constant at higher temperature. Without this term the electrostatic interactions may be overestimated leading to unrealistic protein conformations. In addition to introducing this temperature dependent architecture, we keep
the bare charge of the amino acid at two ends the same.
%%%%%%%%%%%%%%%%%%%%%%%%%%%%%%%%%%%%%%%%%%%%%%%%
\subsection{Structure-based model for the Histone H1 globular domain}

Histone H1 contains a folded globular domain flanked by intrinsically
disordered N- and C-terminal regions. The disordered regions are
described by the same residue-level interaction model used for
ProT$\alpha$, whereas the globular domain is stabilized by a
structure-based native-contact potential constructed from a reference
structure of Histone H1.

Native contacts were identified from the reference structure for
residue pairs separated by at least three positions along the
sequence. A pair $(i,j)$ was classified as a native contact when at
least one pair of side-chain atoms was separated by $4.5$~\AA\ or
less. For each native pair, the reference distance $r_{ij}^{0}$ was
taken to be the corresponding C$_\alpha$--C$_\alpha$ separation.
The contact strength was residue dependent and obtained from the
Miyazawa--Jernigan interaction matrix according to
\begin{equation}
 e_{ij}=\frac{M_{ij}}{2\langle M\rangle},
\end{equation}
where $M_{ij}$ is the matrix element for the corresponding residue
pair and $\langle M\rangle$ is the average matrix value.
Backbone hydrogen-bond contributions identified from the reference
structure were added to the native-contact energy when present.

Native contacts were represented by pair-specific Lennard-Jones
interactions,
\begin{equation}
 U_{ij}^{\rm nat}(r)
 =4\epsilon_{ij}
 \left[
 \left(\frac{\sigma_{ij}}{r}\right)^{12}
 -
 \left(\frac{\sigma_{ij}}{r}\right)^6
 \right],
\end{equation}
where
\begin{equation}
 \sigma_{ij}=\frac{r_{ij}^{0}}{2^{1/6}},
 \qquad
 \epsilon_{ij}=-g_{\rm Go}e_{ij},
\end{equation}
and $g_{\rm Go}=10$ is the native-contact energy scaling factor.
Thus, the minimum of each native-contact potential occurs at its
reference C$_\alpha$--C$_\alpha$ separation. The resulting topology
contains 151 native-contact pairs in the structured region of
Histone H1.

Local structure within the folded region was further stabilized by
harmonic angle potentials whose equilibrium angles were taken from
the reference structure, with
$k_\theta=600$~kJ~mol$^{-1}$~rad$^{-2}$. The globular domain was
therefore not treated as a rigid body; rather, its native architecture
was maintained by the structure-based contact network and local
backbone potentials while retaining conformational fluctuations.

%%%%%%%%%%%%%%%%%%%%%%%%%%%%%%%%%%%%%%%%%%%%%%%%%

%%%%%%%%%%%%%%%%%%%%%%%%%%%%%%%%%%%%%%%%%%%%%%%%%%%%%% 
A known limitation of residue-level coarse-grained models with implicit
solvent and screened Coulomb interactions is the systematic overestimation
of $R_g$ for highly charged IDPs, owing to the neglect of counterion
condensation and Manning charge renormalization~\cite{dignon2018sequence,tesei2021accurate}.
For ProT$\alpha$ specifically, comparison with single-molecule FRET
measurements on isolated chain fragments quantifies this deviation
directly: for the N-terminal fragment (Q~$= -43$) the model yields
$R_g^{\rm sim} = 4.62$~nm versus $R_g^{\rm expt} = 3.78$~nm (22\%
overestimation), and for the C-terminal fragment (Q~$= -40$) we obtain
$R_g^{\rm sim} = 3.93$~nm versus $R_g^{\rm expt} = 3.63$~nm (8\%
overestimation)~\cite{seth2024fine}.
The trend is consistent with a Manning condensation picture: the
overestimation grows with net charge magnitude, as the effective charge
renormalization neglected by the model becomes increasingly important
at higher linear charge density.
 
In the condensate studied here, this limitation is partially mitigated
by two effects specific to the many-chain environment. First, the
near charge-neutrality of the 50 ProT$\alpha$ + 40 Histone H1 mixture
($\Delta Q \approx 50\times(-44) + 40\times(+53) = -80\,e$, nearly
neutral overall) provides many-body internal screening that is absent
for an isolated chain in dilute solution: the field of each ProT$\alpha$
chain is partially compensated by neighboring H1 chains and vice versa.
Second, the inter-chain ProT$\alpha$--H1 electrostatic attraction that
drives condensate formation is also overestimated in the unscreened
limit, partially compensating the overestimated intra-chain
self-repulsion. These two effects act in concert to reduce the
sensitivity of condensate chain dimensions and intermolecular
organization to the implicit-solvent approximation relative to the
dilute-solution case. The chain dimensions reported in Section~III\,C
are therefore expected to carry a smaller relative error than the
dilute-solution values quoted above, though a quantitative assessment
requires explicit-counterion simulations, which remain an important
direction for future work.
\section{Results}
\subsection{ProT$\alpha$--Histone System}
The simulated system consisted of 50 ProT$\alpha$ molecules and 40
histone H1 molecules, resulting in an approximately charge-neutral
mixture. The globular domain of H1 was represented using a
structure-based Go-type native contact model, while the intrinsically disordered regions
of H1 and the entire ProT$\alpha$ chains were modeled using the HPS
coarse-grained model described above.

Initial configurations were generated by placing all molecules at
random non-overlapping positions within a cubic simulation box with
periodic boundary conditions. The resulting configurations were first
energy minimized to remove residual steric overlaps and unfavorable
contacts. All simulations were performed using GROMACS employing a
stochastic dynamics (Langevin) integrator at $T=300$ K with a friction
coefficient of $1~\mathrm{ps}^{-1}$ and a time step of 10 fs.

For each target pressure in the range P = 2–12 bar, the system was
equilibrated in the NPT ensemble for 200 ns using isotropic Berendsen
pressure coupling with a relaxation time of 2 ps. During this stage,
the simulation box volume was allowed to fluctuate until the
equilibrium density corresponding to the imposed pressure was reached.
The average box length obtained from the NPT stage was then used to
define the fixed simulation volume for subsequent NVT production runs.

We emphasize that the pressure range studied here (2–12 bar) is not
intended to model any physical compression scenario relevant to
biology — structural transitions in proteins require pressures of
order kilobars, and chain conformations are indeed insensitive to
compression over the range studied (Section III C). Rather, pressure
serves as a precise and reversible thermodynamic handle to tune the
condensate packing fraction $\phi$ while holding temperature, chain length,
and all interaction parameters fixed.

The equilibrium box lengths decrease monotonically from L=22.98~nm at P=2~bar to L=17.40~nm at 
P=12~bar. The corresponding volume
fractions, estimated as $\phi = N_{total} V_{bead} / V_{box}$ using
the effective bead volume $V_{bead} \approx \ell_0^3$, where $\ell_0 =
0.38 $ nm is the C$_\alpha$--C$_\alpha$ bond length and excluded volume
interactions among bonded neighbors are absent in the model,
span $\phi \approx 0.06–0.14$ across the pressure range, consistent with the dilute-to-semidilute regime.
The results should therefore be interpreted as
characterizing how condensate structure, dynamics, and intermolecular
organization evolve with increasing chain packing fraction.

\begin{figure}[t]
  \centering 
  \includegraphics[width=0.4\textwidth]{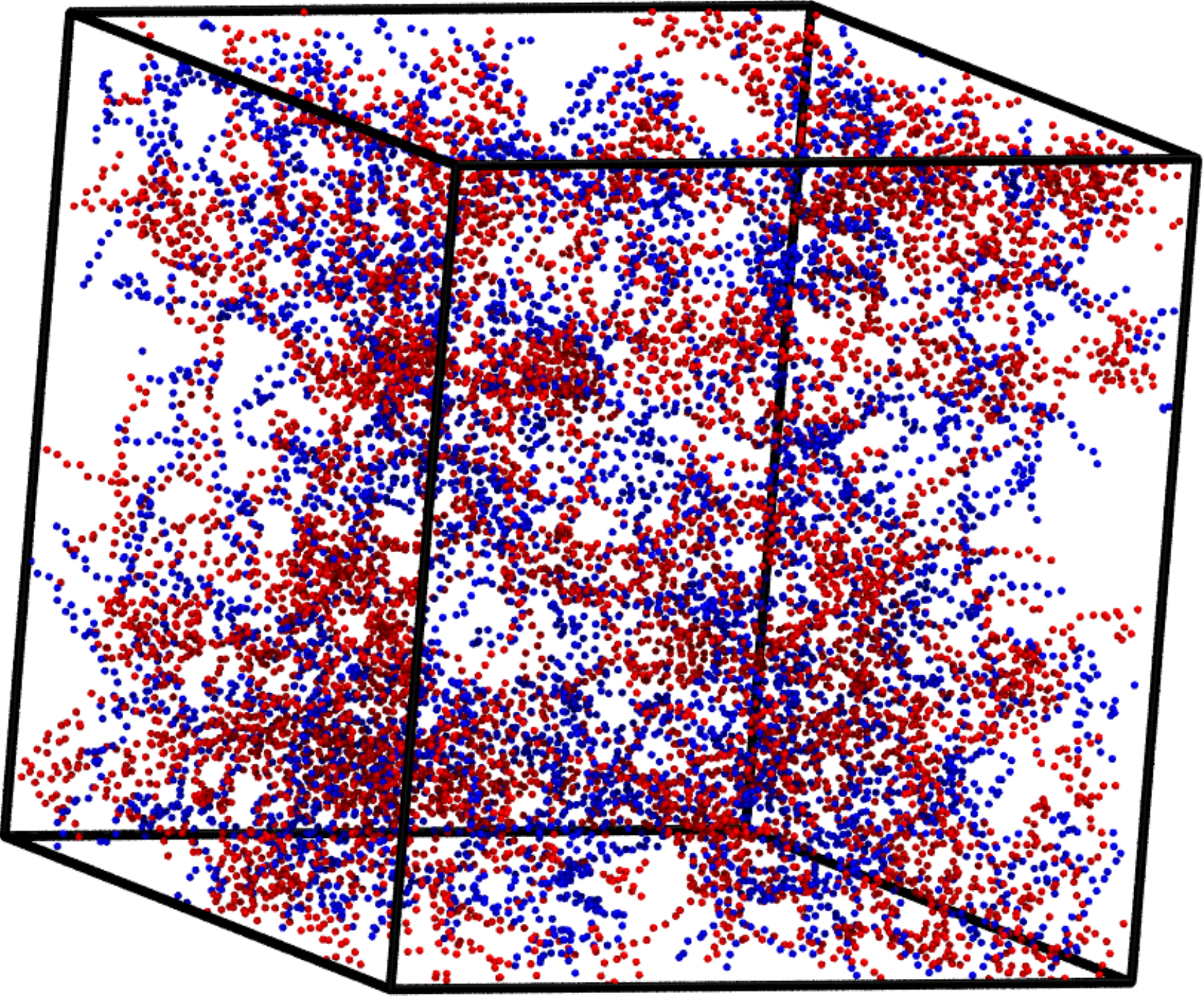}
  \vspace{0.25truecm}
  \includegraphics[width=0.4\textwidth]{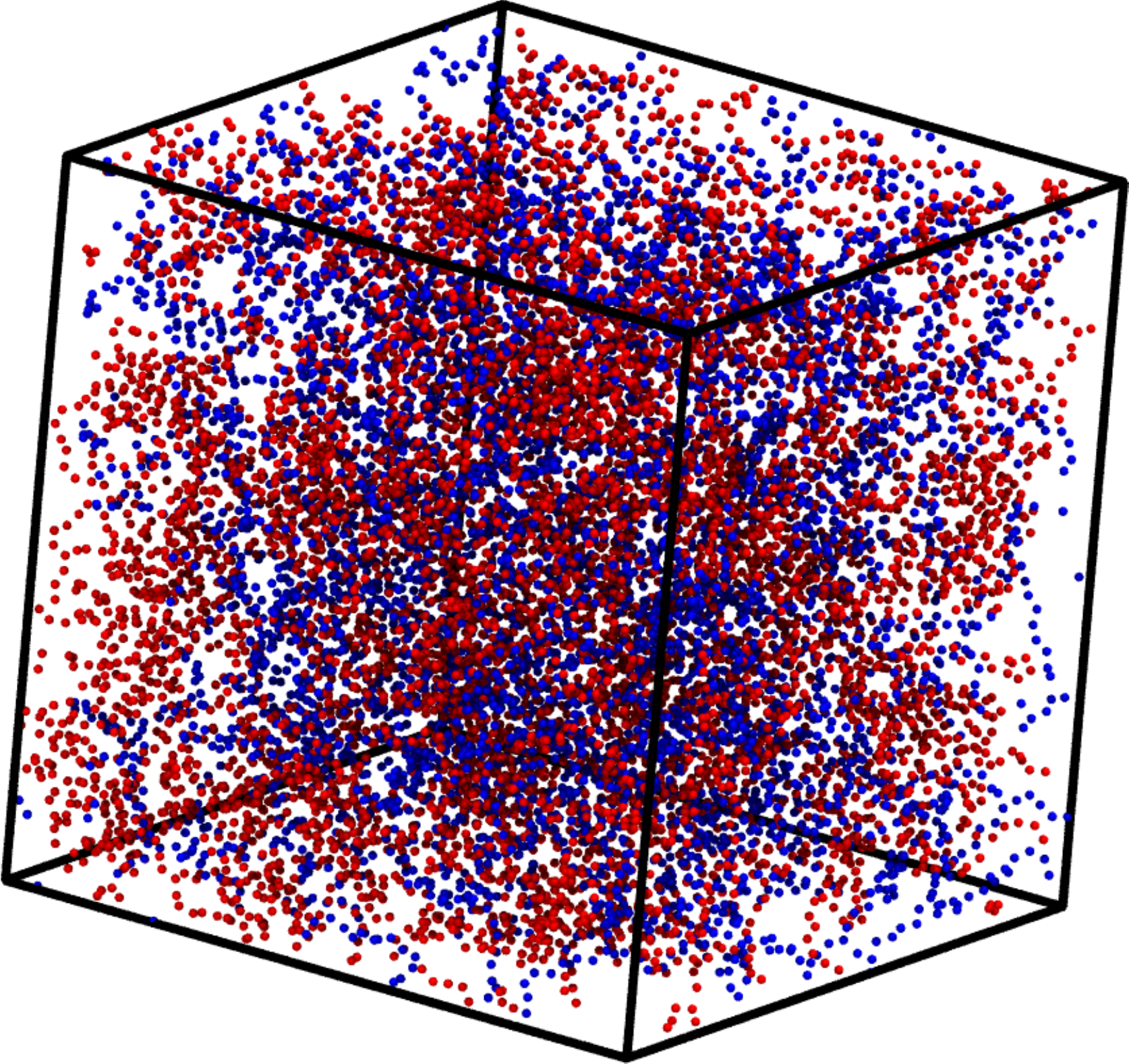}
  \caption{\small Representative configurations of the ProT$\alpha$–H1
    mixture at (top) \(P=2\) bar (\(L=22.98\) nm) and (bottom) \(P=12\) bar
    (\(L=17.40\) nm). ProT$\alpha$ and Histone H1 residues are shown
    in blue and red, respectively. Each system contains 50
    ProT$\alpha$ and 40 Histone H1 molecules. Molecules are shown
    within the primary periodic simulation cell. The panels are
    displayed at the same graphical size for clarity; the indicated
    box dimensions reflect the pressure-dependent compression of the
    system. Periodic boundary condition is applied to generate the figures using
    VMD~\cite{humphrey1996vmd}.}
  \label{fig:snapshots}
\end{figure}

Production simulations were carried out in the $NVT$ ensemble for an
additional 100 ns with the volume fixed at its equilibrium value and
the pressure coupling turned off. Configurations were saved every
0.1 ns, yielding 1000 snapshots for structural and thermodynamic
analysis. Electrostatic and hydrophobic interactions were truncated at
2.5 nm, and periodic boundary conditions were applied in all three
spatial directions. Typical snapshots for frame 500 is shown in Fig.~\ref{fig:snapshots}.
%%%%%%%%%%%%%%%%%%%%%%%%%%%%%%%%%%%%%%%%%%%%%%%%%
%%%%%%%%%%%%%%%%%%%%%%%%%%%%%%%%%%%%%%%%%%%%%%%%%%%
%%%%%%%%%%%%%%%%%%%%%%%%%%%%%%%%%%%%%%%%%%%%%%%%%%%%%%%%%%%%
\subsection{Structural and Dynamical Measurements}
Steady-state structural properties were obtained by averaging over the
$N_{\rm frame}=1000$ configurations stored during the production
trajectory. For a generic observable $\Psi$, the ensemble average was
computed as

\begin{equation}
\langle \Psi \rangle = \frac{1}{N}\sum_{i=1}^{N_{\rm frame} }\Psi_i
\end{equation}
where $\Psi_i$ denotes the value of the observable in the $i^{\rm th}$
saved configuration. Quantities evaluated in this manner include the
mean-square radius of gyration $\langle R_g^2\rangle$, the mean-square
end-to-end distance $\langle R_N^2\rangle$, their ratio
$R=\langle R_N^2\rangle/\langle R_g^2\rangle$ (Section III C), and the
radial distribution functions of both individual coarse-grained beads
and chain centers of mass (Sections III D--F).

Dynamical properties were obtained from time-correlation functions
averaged over all possible time origins along the trajectory. For a
time-dependent vector quantity $\vec{\zeta}(t)$, we define
\begin{equation}
C(\Delta t)= \left\langle \vec{\zeta}(t) \cdot \vec{\zeta}(t+\Delta
  t)\right\rangle_t
\label{eq:correlation}
\end{equation}
where $\langle \cdots \rangle_t$ denotes an average over all
time origins $t$. The same procedure was used to compute chain
relaxation functions (Section III I) and the mean-square displacement
(MSD) (Section III H), from which translational diffusion coefficients
were obtained.
%%%%%%%%%%%%%%%%%%%% Rg-RDF section %%%%%%%%%%%%%%%%%%%%%%%
%%%%%%%%%%%%%%%%%%%%%%%%%%%%%%%%%%%%%%%%%%%%%%%%%%%%%%%%%%%% beginning of Gyration Radii 
\subsection{Gyration Radii}
\label{sec:Rg}
While presenting the results for the PH system it is worthwhile to note
that for a fully flexible Gaussian chain the distributions of end-to-end distance
$R_N \equiv \sqrt{\langle R_N^2\rangle}$ and the gyration
radius $R_g \equiv \sqrt{\langle R_g^2\rangle}$ are Gaussian, and the
ratio $R = \langle R_N^2\rangle/\langle R_g^2\rangle =
6.0$~\cite{rubinstein2003polymer}.
%%%%%%%%%%%%%%%%%%% Rg-RDf.tex %%%%%%%%%%%%%%%%%%%%%%%%%%%%%%%%
\begin{figure*}[ht!]
  \centering 
  \includegraphics[width=0.95\textwidth]{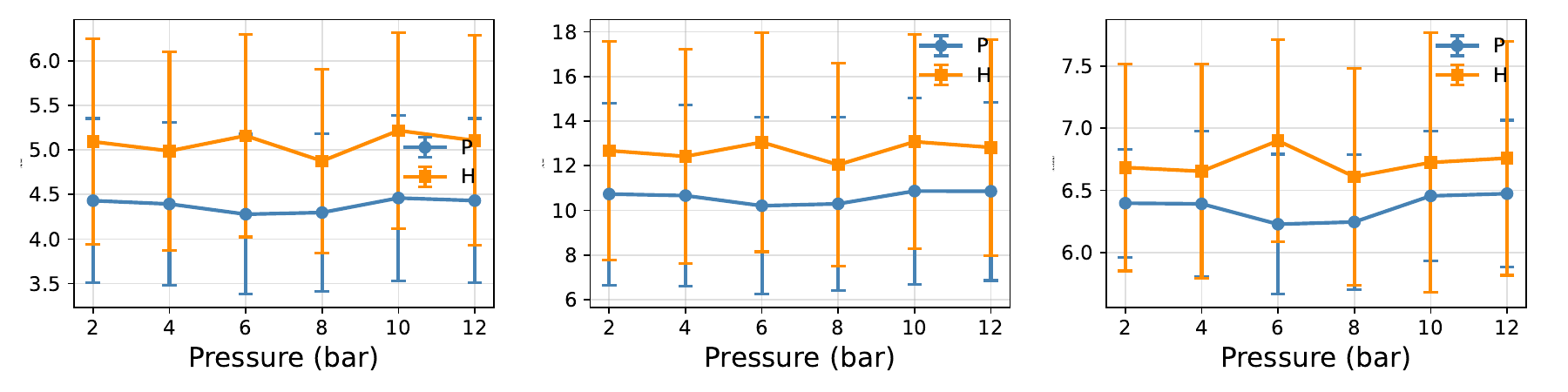}
  \includegraphics[width=0.95\textwidth]{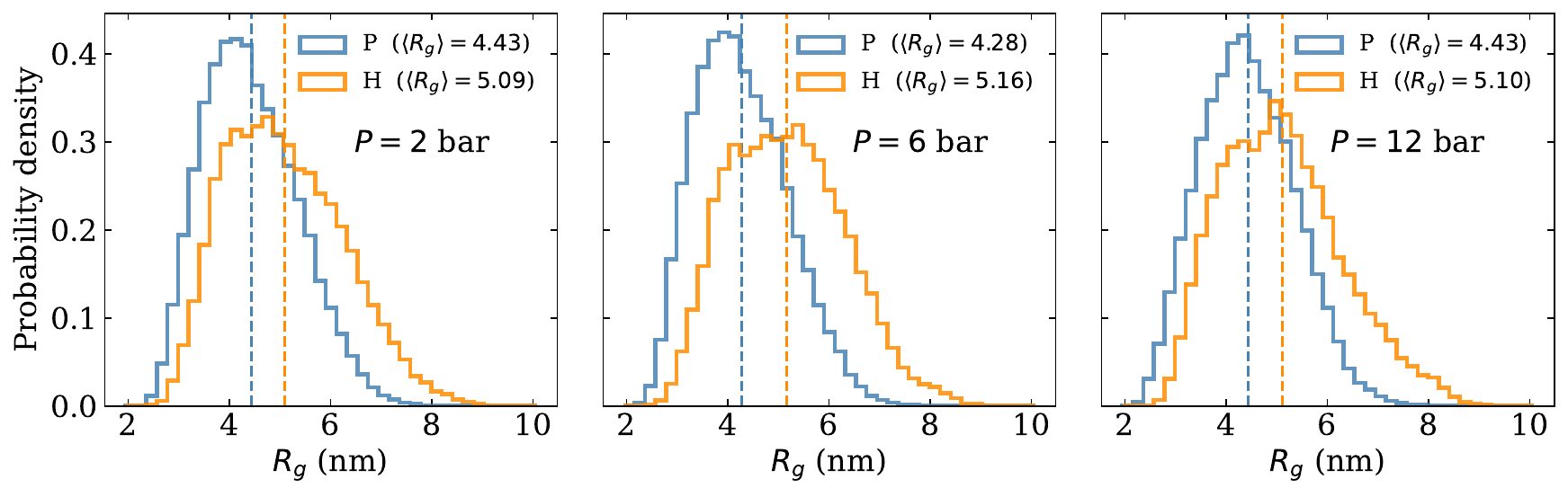}
  \vspace{-6pt}
\caption{\small (top row)
Structural properties of ProT$\alpha$ (P) and Histone H1 (H) as a
function of pressure. Shown are the average radius of gyration
$\langle R_g\rangle$ (left), end-to-end distance $\langle R_n\rangle$
(middle), and the conformational ratio $R=\langle R_n^2\rangle/\langle
R_g^2\rangle$ (right). All three quantities exhibit only weak pressure
dependence over the range 2--12 bar. Histone H1 displays
systematically larger values of $\langle R_g\rangle$, $\langle
R_n\rangle$, and $R$ than ProT$\alpha$, reflecting its
globular-core--flexible-tail architecture. Error bars denote standard
deviations.
(Bottom row) Probability distributions of the radius of gyration ($R_g$), for ProT$\alpha$ (P) and Histone H1 (H) at representative pressures $P=2$, 6, and 10 bar. Only minor changes are observed with pressure. Histone H1 exhibits systematically larger values and broader distributions than ProT$\alpha$, consistent with its globular-core--flexible-tail architecture.
}
\label{fig:rg-versus-p}
\vspace{-4.0pt}
\end{figure*}
%%%%%%%%%%%%%%%%%%%%%%%%%%%%%%%%%%%%%%%%%%%%%%%%%%%%%%%%%%%%
In our PH system the ProThymosin chains are modeled as fully flexible
chains with a relatively high net negative charge, while Histone has
globular domains with two fully disordered ends and a net
positive charge. Excluded volume (EV)
and electrostatic repulsion are responsible for swelling of the chain ($R > 6$),
while hydrophobic contacts and chain stiffness are responsible for
chain compactification ($R < 6$).
Since the present simulations contain no explicit counterions, counterion condensation and ion-mediated correlations are not represented explicitly, although electrostatic interactions are screened through the Debye--H\"uckel term.
%Since the present simulations
%contain no explicit counter-ions, electrostatic interactions are
%fully unscreened and no counterion condensation operates.
%%%%%%%%%%%%%%%%%%%%%%%%%%%%%%%%%%%%%%%%%%%%%%%%%%%%%%%%%%%%
We observe that the gyration radii, end-to-end distances, and the
corresponding ratio $R$ at
different pressure do not show significant variation within error
bars as shown in Fig.~\ref{fig:rg-versus-p} (top row). However, the ratio $R$ is systematically
larger for Histone ($R = 6.7$--$6.9$) than for ProT$\alpha$
($R = 6.2$--$6.4$), a noticeable departure from Gaussian statistics.
The pressure independence within such a narrow window $P=2$--$12$~bar is
expected, as it requires several kilobars for proteins to undergo structural
transitions. However, the
chain conformations within the condensate leading to distributions of
these quantities are worth further analysis Fig.~\ref{fig:rg-versus-p}
(botom row). Histone has a compact globular core and two disordered charged tails
(N-terminal and C-terminal). The globular core compacts the mass
distribution leading to moderate $R_g$, but the two tails extend outward in
opposite directions resulting in large $R_N$. Thus, the ratio $R$
inflates for H, not because H is more extended overall, but because
its termini are far apart while its core is compact. The near-Gaussian
behavior of P, on the contrary, suggests these effects nearly cancel
— a property of the specific sequence and charge distribution of
ProT$\alpha$ but not a general rule.
For Histone the argument breaks down because the globular domain introduces
a non-random chain architecture: the core-tail structure systematically
inflates $R$ above the Gaussian value regardless of the cancellation
argument.
\subsection{Radial Distribution Functions}
\label{sec:rdf}
We compute two complementary classes of radial distribution functions (RDFs) to characterize
the structural organization of the binary ProT$\alpha$--Histone system across the pressure range
$P = 2$--$12$~bar. The bead-level RDFs $g_{PP}(r)$, $g_{HH}(r)$, and $g_{PH}(r)$ are computed
over all pairs of C$_{\alpha}$ beads belonging to the indicated species, capturing intra-chain geometry and
local packing at single-residue resolution (section ~\ref{sec:bead-rdf}
and Fig.~\ref{fig:bead-rdf}). The center-of-mass (CM) RDFs $g_{PP}^{\rm CM}(r)$,
$g_{HH}^{\rm CM}(r)$, and $g_{PH}^{\rm CM}(r)$ are computed over chain centers of mass,
providing information about intermolecular organization, clustering, and mixing at the
whole-chain level (section~\ref{sec:cm-rdf} and Fig.~\ref{fig:cm-rdf}). Together, the two levels of description yield a holistic structural picture that
neither alone can provide.
%We note that all
%electrostatic interactions between chains are unscreened, which has direct consequences
%for the observed spatial correlations discussed below.
%----------------------------------------------------------------------
\subsection{Bead-Level Radial Distribution Functions}
\label{sec:bead-rdf}
Figure~\ref{fig:bead-rdf} shows the bead-level RDFs at three representative pressures
($P = 2.0$, $6.0$, and $12.0$~bar). A systematic summary of all peak positions and heights
across all pressures for ProT$\alpha$ and Histone are given in
Tables~\ref{tab:pp-peaks} and \ref{tab:hh-peaks} respectively.
\subsubsection{P--P bead-level RDF}
\label{section:PP-bead-rdf}
The bead-level $g_{PP}(r)$ (Fig.~\ref{fig:bead-rdf}(a)) exhibits a series of sharp peaks at $r = 0.70$, $0.90$, $1.04$,
$1.36$, and $1.70$~nm, whose positions are strictly invariant with pressure across the full
range studied. ProT$\alpha$ is an intrinsically disordered protein (IDP) carrying no
Gō-type native-contact potential in the CG model; its chain is governed by bonded interactions
alone — bond length, bond angle, and soft dihedrals. With a C$\alpha$--C$\alpha$ virtual bond
length $\ell_0 \approx 0.38$~nm and an equilibrium bond angle set by the model parameterization,
the expected intra-chain bead separations for the 1--3, 1--4, 1--5, 1--7, and 1--9 pairs
along the backbone are approximately $0.70$, $0.90$, $1.04$, $1.36$,
and $1.70$~nm, respectively, in quantitative agreement with the observed peak positions. These peaks are  
therefore a purely geometric fingerprint of chain connectivity; pressure cannot alter bonded  
distances and the positions are consequently pressure-independent.  
%%%%%%%%%%%%%%%%%%%%%%%%%%%%%%%%%%%%%%%%%%%%%%%%%
\begin{table}[ht!]
  \footnotesize 
  \caption{Peak positions and heights of the P--P bead-level RDF; 
  Pressures are expressed in bar and distances are in nanometers.}
\label{tab:pp-peaks}
\centering 
\begin{tabular}{ccccccccccc}
  \hline\hline 
  $P$ & $r_1$& $g_1$ & $r_2$ & $g_2$ & $r_3$& $g_3$ & $r_4$ & $g_4$ &
                                                                    $r_5$&  $g_5$\\\hline 
2  & 0.70 & 21.81 & 0.90 & 2.06 & 1.04 & 3.65 & 1.36 & 2.02 & 1.70 & 1.37\\
4  & 0.70 & 14.20 & 0.90 & 1.48 & 1.04 & 2.55 & 1.36 & 1.51 & 1.70 & 1.14\\
6  & 0.70 & 12.20 & 0.90 & 1.36 & 1.04 & 2.26 & 1.36 & 1.40 & 1.70 & 1.10\\
8  & 0.70 & 10.94 & 0.90 & 1.28 & 1.04 & 2.10 & 1.36 & 1.34 & 1.70 & 1.08\\
10 & 0.70 & 10.18 & 0.90 & 1.24 & 1.04 & 2.00 & 1.36 & 1.31 & 1.70 & 1.09\\
12 &  0.70 & 9.66 & 0.90 & 1.22 & 1.04 & 1.93 & 1.36 & 1.29 & 1.70 & 1.07\\
\hline\hline 
\end{tabular}
\end{table}
\iffalse
%%%%%%%%%%%%%%%%%%%%%%%%%%%%%%%%%%%%%%%%%%%%%%%%%
%%%%%%%%%%%%%%%%%%%%%%% RDF%%%%%%%%%%%%%%%%%%%%
\begin{figure}[H]
  \centering 
\includegraphics[width=0.9\linewidth]{PDF/Fig3a.pdf}
\includegraphics[width=0.9\linewidth]{PDF/Fig3b.pdf}
\includegraphics[width=0.9\linewidth]{PDF/Fig3c.pdf}
\vspace{-6pt}
\caption{\small Bead-level radial distribution functions 
 (a) $g_{PP}(r)$, (b) $g_{HH}(r)$, and (c) $g_{PH}(r)$) at representative pressures $P=2$ (blue), 6 green), and 12 
  (red) bar. Insets show the short-distance region on an expanded 
  scale. The positions of the dominant peaks are insensitive to 
  pressure, indicating preservation of local molecular structure 
  throughout the compression range. In contrast, peak amplitudes 
  decrease with increasing pressure owing to the increase in overall 
  density (Eqn.~\ref{eq:scaling}). The unlike-species correlation 
  function $g_{PH}(r)$ excepting for P=2.0 bar remains close to unity beyond the excluded-volume regime, consistent with a well-mixed condensate.}
\label{fig:bead-rdf}
\end{figure}
%%%%%%%%%%%%%%%%%%%%%%%%%%%%%%%%%%%%%%%%%%%%%%%%%
\fi
The peak heights decrease monotonically with increasing pressure. Strikingly, all five peaks
are suppressed  by the same universal factor: between $P = 2.0$~bar and $P = 12.0$~bar the ratio is $g_{\rm peak}(2\,\text{bar})/g_{\rm peak}(12\,\text{bar}) \approx 2.26$ for
every peak. This is quantitatively accounted for by the volume compression ratio alone,
\begin{equation}
\begin{aligned}
\frac{g_{\rm peak}(P_1)}{g_{\rm peak}(P_2)}
\approx \frac{V(P_1)}{V(P_2)} 
&= \left(\frac{L(P_1)}{L(P_2)}\right)^3 \\
&= \left(\frac{22.982}{17.402}\right)^3 \approx 2.29
\end{aligned}
\label{eq:scaling}
\end{equation}
in excellent agreement with observation. The physical mechanism is the following. At short $r$,
$g_{PP}(r)$ is dominated by intra-chain bead pairs: owing to the strong electrostatic
repulsion between like-charged ProT$\alpha$ chains, inter-chain
contributions at short distances are strongly suppressed. The number of intra-chain pairs at
each separation distance is fixed by chain topology and independent of pressure. As the box
contracts, the mean P-bead number density $\rho_P = N_P/V$ increases, raising the normalization
denominator of $g(r)$ while the numerator remains essentially constant, yielding
$g_{\rm peak} \propto 1/\rho_P \propto V(P)$. This is confirmed independently by the
radius of gyration data (Fig.~\ref{fig:rg-versus-p}): $\langle R_g^P \rangle \approx 4.8$~nm is
pressure-independent within statistical uncertainty, establishing that ProT$\alpha$ undergoes
no pressure-induced compaction over this range. The ratio
$\langle R_n^2\rangle/\langle R_g^2\rangle \approx 6.4$ is consistent with the
Gaussian-chain theoretical value of $6$~\cite{rubinstein2003polymer}, confirming the IDP
character of ProT$\alpha$ in the CG representation.

The universal suppression factor — identical for all five peaks — excludes any
selective disruption of particular bead-pair separations. No pressure-induced unfolding,
conformational transition, or structural reorganization occurs in ProT$\alpha$ across the
pressure range studied.

%%%%%%%%%%%%%%%%%%%%%%%%%%%%%%%%%%%%%%%%%%%%%%%%%
\begin{table}[ht!]
  \footnotesize 
\caption{Peak positions and heights of the H--H bead-level RDF; 
  Pressures are expressed in bar and distanes are in nanometer}
\label{tab:hh-peaks}
\centering 
\begin{tabular}{ccccccccccc}
\hline\hline 
$P$ & $r_1$& $g_1$ & $r_2$ & $g_2$ & $r_3$& $g_3$ & $r_4$ & $g_4$ &
                                                                    $r_5$
  &  $g_5$\\
\hline 
2  & 0.54 & 5.73 & 0.70 & 11.64 &0.88 & 2.64& 1.02 & 3.76& 1.36 & 2.60\\
4  & 0.54 &  3.76 &0.70 & 7.62 & 0.88 & 1.81&1.02 & 2.56&1.36 & 1.83\\
6  & 0.54 &  3.20 &0.70 & 6.50 & 0.88 & 1.60&1.02 & 2.24&1.36 & 1.65\\
8  & 0.54 &  2.87 &0.70 & 5.86 & 0.88 & 1.47&1.02 & 2.06&1.36 & 1.54\\
10 & 0.54 &  2.69 &0.70 & 5.45 & 0.88 & 1.41&1.02 & 1.96&1.36 & 1.49\\
12 & 0.54 &  2.53 &0.70 & 5.18 & 0.88 & 1.35&1.02 & 1.88&1.36 & 1.44\\
\hline\hline 
\end{tabular}
\end{table}
%%%%%%%%%%%%%%%%%%%%%%%%%%%%%%%%%%%%%%%%%%%%%%%%%%%%%%%%%%%%%%%%%%%%%
\subsubsection{H--H bead-level RDF}
\label{section:HH-bead-rdf}
The bead-level $g_{HH}(r)$ (Fig.~\ref{fig:bead-rdf}(b)) displays peaks at $r = 0.54$, $0.70$, $0.88$, $1.02$, and
$1.36$~nm, again with strictly invariant positions. Two distinct contributions are present.
The first is the same bonded-geometry origin as in ProT$\alpha$: backbone bead separations
imposed by the CG bond and angle terms. The second, specific to Histone, is the set of native
contact distances encoded in the Gō-type native contact potential, which enforces bead pairs at their
crystallographic separations from the folded PDB structure. Since neither bonded parameters
nor Gō-type native contact distances are altered by the applied pressure, the peak positions are
pressure-independent, now for a richer physical reason than in ProT$\alpha$.
%%%%%%%%%%%%%%%%%%%%%%%%%%%%%%%%%%%%%%%%%%%%%%%%%
%%%%%%%%%%%%%%%%%%%%%%% RDF%%%%%%%%%%%%%%%%%%%%
\begin{figure}[ht!]
  \centering 
\includegraphics[width=0.9\linewidth]{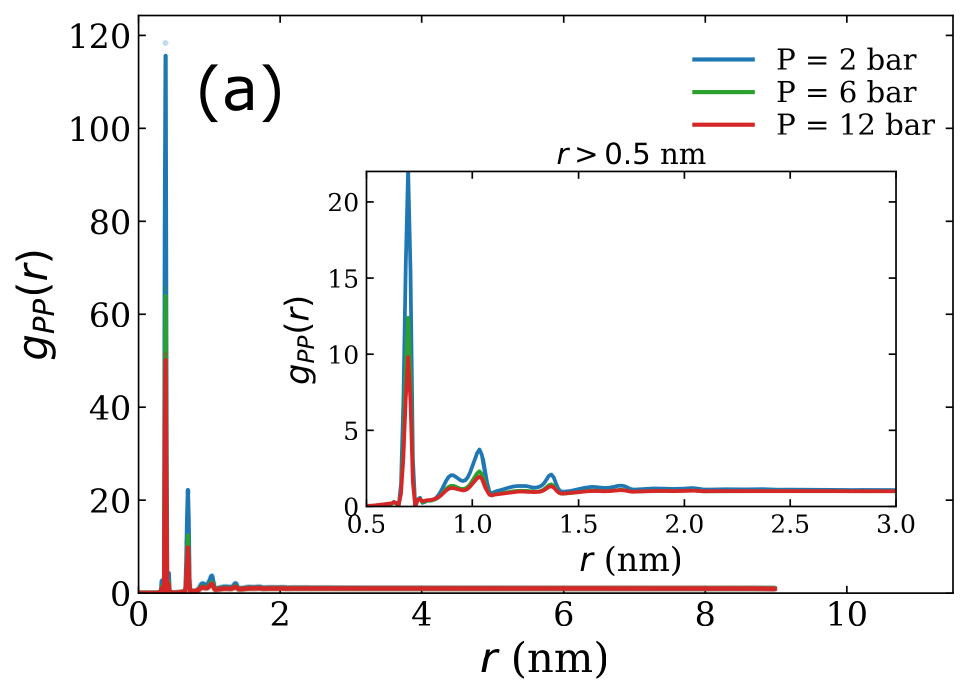}
\includegraphics[width=0.9\linewidth]{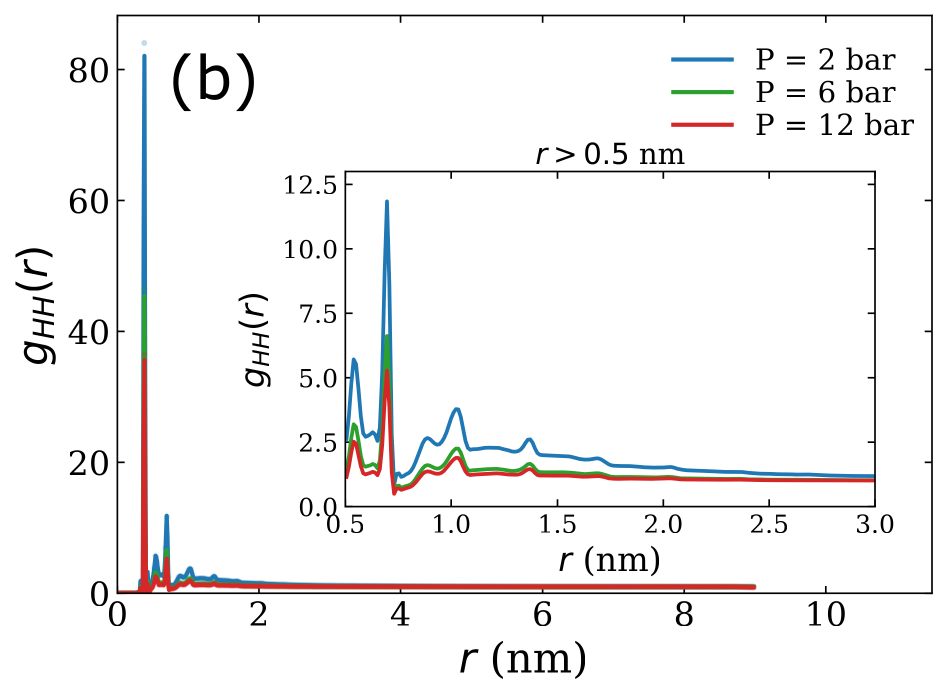}
\includegraphics[width=0.9\linewidth]{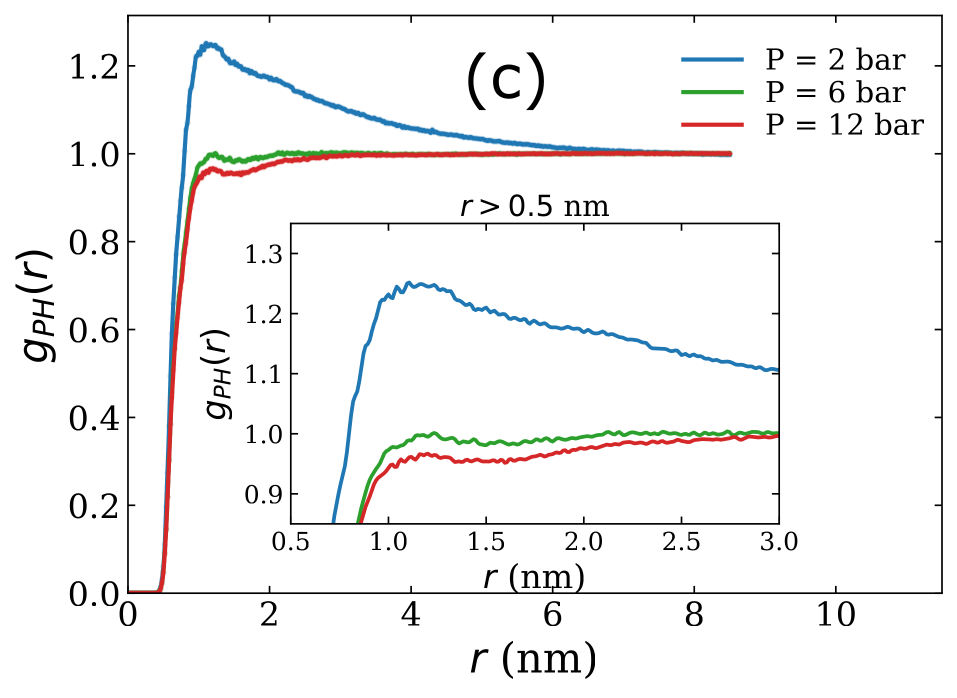}
\vspace{-6pt}
\caption{\small Bead-level radial distribution functions 
 (a) $g_{PP}(r)$, (b) $g_{HH}(r)$, and (c) $g_{PH}(r)$) at representative pressures $P=2$ (blue), 6 green), and 12 
  (red) bar. Insets show the short-distance region on an expanded 
  scale. The positions of the dominant peaks are insensitive to 
  pressure, indicating preservation of local molecular structure 
  throughout the compression range. In contrast, peak amplitudes 
  decrease with increasing pressure owing to the increase in overall 
  density (Eqn.~\ref{eq:scaling}). The unlike-species correlation 
  function $g_{PH}(r)$ excepting for P=2.0 bar remains close to unity beyond the excluded-volume regime, consistent with a well-mixed condensate.}
\label{fig:bead-rdf}
\end{figure}
%%%%%%%%%%%%%%%%%%%%%%%%%%%%%%%%%%%%%%%%%%%%%%%%%

The peak heights are again universally suppressed with pressure by the same factor
$V(P_1)/V(P_2)$ as observed for P--P (Table~\ref{tab:pp-peaks}). This identity of the
suppression factor across two structurally distinct chain models — a fully disordered IDP
and a native-contact-stabilized globular domain — confirms that the bead-level RDF at short distances is
dominated by intra-chain pairs in both cases. The intermolecular clustering of Histone
chains, clearly visible in the CM-RDF (Sec.~\ref{sec:cm-rdf}), contributes only a minor
additive correction to the total bead-level signal and does not break the universal
$g_{\rm peak} \propto V(P)$ scaling.

This is corroborated by the radius of gyration: $\langle R_g^H \rangle \approx 5.0$~nm
is pressure-independent Fig.~\ref{fig:rg-versus-p}), confirming that
the native-contact-stabilized  Histone
fold is mechanically robust across the full compression range and does not unfold under
the applied pressures. The ratio $\langle R_n^2\rangle / \langle R_g^2\rangle \approx 6.7$
for Histone, slightly above the Gaussian value, reflects the additional stiffness
introduced by the native-contact network.

%..............................................
\subsubsection{P--H bead-level RDF}
\label{section:PH-bead-rdf}
The P--H bead-level RDF~(Fig.~\ref{fig:bead-rdf}(c)) occupies a qualitatively distinct category from $g_{PP}(r)$ and $g_{HH}(r)$ because, by construction, all P--H bead pairs are inter-chain. There are no intra-chain P--H pairs, no bonded-geometry peaks, and no G={o}-model native-contact peaks to dominate the signal. Consequently, $g_{PH}(r)$ directly probes inter-species spatial correlations at the bead level.

At the lowest pressure ($P=2$ bar), $g_{PH}(r)$ exhibits a broad
maximum at $r \approx 1.1$ nm with $g_{PH}(r)\approx 1.25$, indicating a modest enhancement of local P--H
contacts relative to random mixing. With increasing pressure this
low-$r$ enhancement is suppressed, and the RDF develops a shallow
minimum near $r\approx 1.5$ nm, most clearly visible at $P=6$ and 12
bar in the enlarged insets of Fig.~4. Thus, the local P--H bead-level
correlation changes from weakly attractive at low pressure to weakly
depleted at intermediate separations under compression.
We attribute this nonmonotonic behavior primarily to crowding and excluded-volume packing, which increasingly mask the specific
electrostatic P--H contact signal in the ensemble-averaged bead-level
RDF. At larger distances, $g_{PH}(r)$ approaches unity, indicating
that the system remains well mixed at the bead level without
long-range P--H segregation.

This bead-level picture should be interpreted together with the
center-of-mass RDFs discussed below. Although the specific low-$r$
bead-level P--H peak is weakened under compression, the corresponding
CM-RDF shows enhanced probability at short center-of-mass separations,
demonstrating substantial interpenetration of P and H chains at the
whole-chain level.

Thus, the bead- and CM-level RDFs provide complementary information:
local bead-level contacts are weak and transient, whereas whole-chain
correlations reveal persistent P--H mixing and interpenetration driven
by charge complementarity. Individual P--H contacts continuously
exchange on sub-nanosecond timescales (please see section~\ref{sec:relax}), while the chains remain associated over much longer relaxation times, consistent with the dynamic and heterogeneous interactions expected for a fuzzy complex \cite{borgia2018extreme,chowdhury2023driving}.
%----------------------------------------------------------------------
\subsection{Center-of-Mass Radial Distribution Functions}
\label{sec:cm-rdf}
Figure~\ref{fig:cm-rdf} shows the CM-level RDFs at $P = 2.0$, $6.0$,
and $12.0$~bar. Together with the bead level RDFs they provide a
complete picture of the intra- and inter-chain organization for 
both ProT$\alpha$ and Histone as well as transient P--H complexes
which will further be probed in the next section. 
Because the CM-RDF integrates over all bead-level correlations within each chain, the
intra-chain bonded-geometry peaks that dominate $g_{PP}(r)$ and $g_{HH}(r)$ at the bead
level are entirely absent here but provide a cleaner probe of
\emph{inter-molecular} spatial organization.
%%%%%%%%%%%%%%%%%%%%%%%%%%%%%%%%%%%%%%%%%%%%%%%%%%%%%%%%%%%%%%
\subsubsection{P--P CM-RDF: $g_{PP}^{CM}(r)$}
\label{section:PP-CM-rdf}
The CM-RDF $g_{PP}^{CM}(r)$ for ProT$\alpha$ 
shows depletion at short distances $r < 1$~nm due to electrostatic
repulsion among ProT$\alpha$ chains, and saturates to unity at $r
\approx 6$~nm (Fig.~\ref{fig:cm-rdf}(a)). This phenomenon is
relatively straightforward to understand. The highly negatively
charged and fully disordered ProT$\alpha$ chains at low pressure have
sufficient free volume to avoid one another due to the 
electrostatic repulsion, resulting in
$g_{PP}^{CM}(r) < 1$ at short distances. The rise to the saturation
value at $r \approx 6$~nm is relatively insensitive to pressure. The average $\langle
R_g^P \rangle \approx 4.3$~nm; if the system consisted of ProThymosin
only, saturation would occur at $\approx 2\langle
R_g^P \rangle \simeq 8.6$~nm. However, the presence of highly positively
charged Histone chains that interpenetrate ProT$\alpha$
brings ProT$\alpha$ chains closer to each other, thereby
reducing this saturation distance to $\approx 6$~nm, as is evident
from $g_{PH}^{CM}(r)$ (Fig.~\ref{fig:cm-rdf}(c)).
This can be viewed as a 
mean-field many-body effect in the bulk system.
%%%%%%%%%%%%%%%%%%%%%%%%%%%%%%%%%%%%%%%%%%%%%%%%%%%%
\subsubsection{H--H CM-RDF: $g_{HH}^{CM}(r)$}
\label{section:HH-CM-rdf}
\begin{figure}[H]
\centering 
\includegraphics[width=0.90\linewidth]{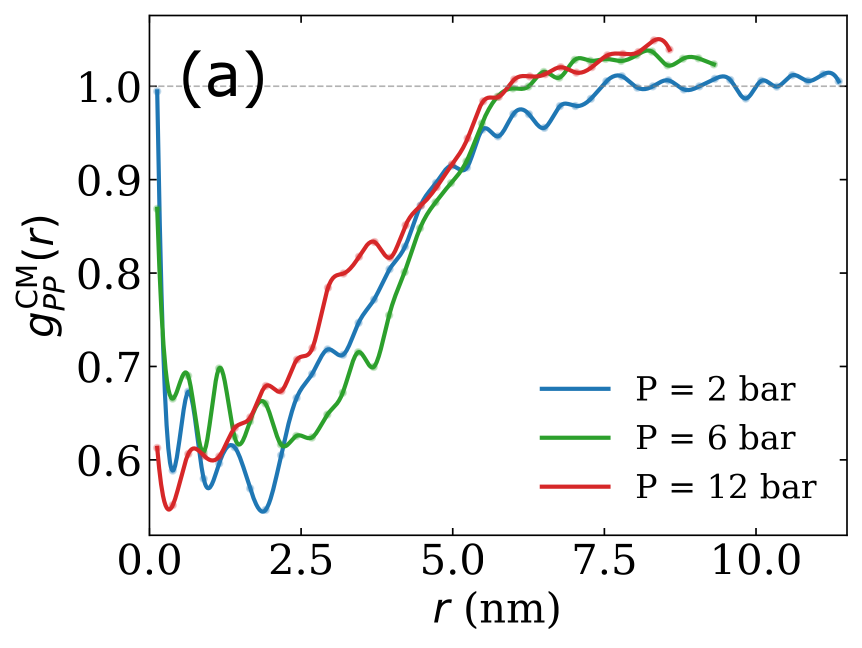}
\includegraphics[width=0.90\linewidth]{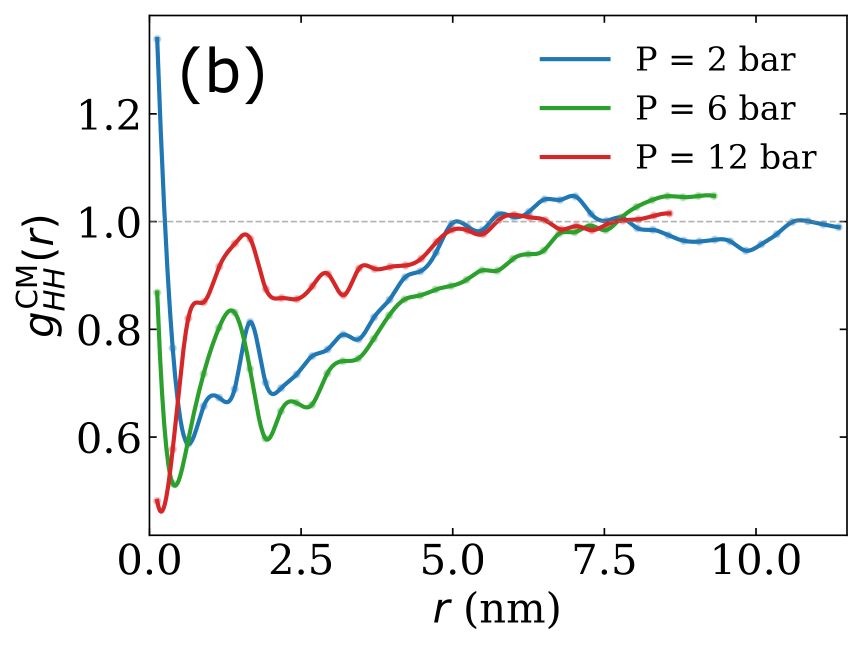}
\includegraphics[width=0.90\linewidth]{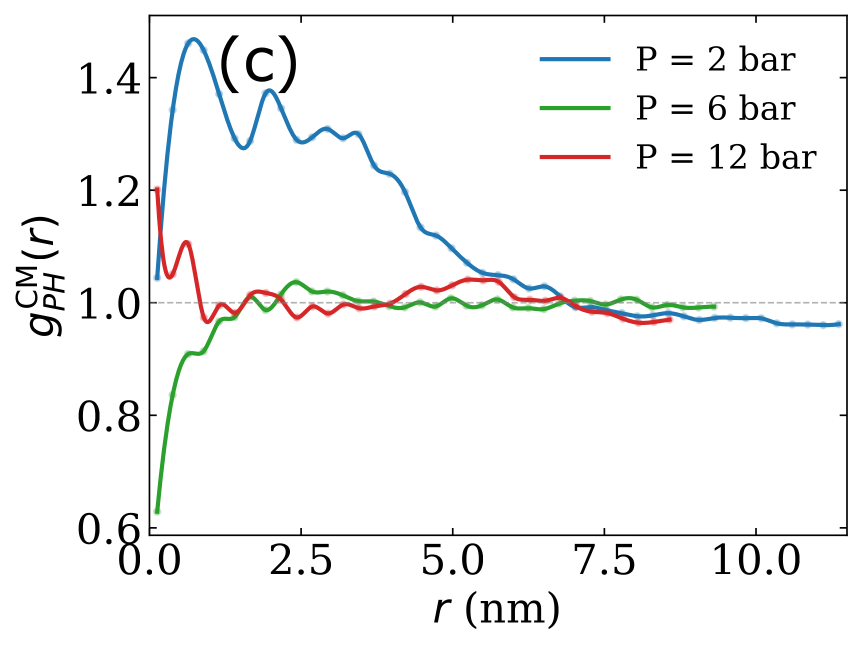}
\vspace{-8pt}
\caption{\small Center-of-mass radial distribution functions 
  $g_{PP}(r)$ (a), $g_{HH}(r)$ (b), and $g_{PH}(r)$ (c) at representative 
  pressures $P=2$ (blue), 6 (green), and 12 (red) bar. Unlike the bead-level RDFs of 
  Fig.~\ref{fig:bead-rdf}, the center-of-mass correlations directly probe intermolecular organization.}
\label{fig:cm-rdf} 
\end{figure}

$g_{HH}^{CM}(r)$ also shows depletion at short distances, however
exhibits a clear peak at $r \approx 1.8$ nm, indicating
net attractive inter-chain correlations and clustering of
Histone chains (Fig.~\ref{fig:cm-rdf}(b)). Unlike ProT$\alpha$, where the only dominant mechanism is
electrostatic interaction among the P-P chains, the H-H 
clustering is driven by the electrostatic, hydrophobic and native-contact interactions encoded in the G$\bar{o}$
potential between folded Histone domains: the G$\bar{o}$ attractive
wells are possibly sufficient to
overcome any like-charge electrostatic repulsion between the basic
Histone chains

It is worthwhile to note that the peak position remains approximately
fixed at $r \approx 1.8$ nm throughout the entire pressure range,
indicating that the characteristic
H-H separation is largely insensitive to compression. However, the
peak becomes more pronounced and broader
at intermediate and high pressures, while the surrounding region
becomes increasingly depleted relative to random mixing.
%%%%%%%%%%%%%%%%%%%%%%%%%%%%%%%%%%%%%%%%%%%%%%%%%
These features suggest enhanced local H-H correlations under
compression without a significant change in the preferred inter-chain
separation.
At $P=2$ bar, $g_{HH}^{CM}(r)$ approaches unity only at $r \approx
8$~nm, whereas at P=12 bar the same saturation occurs near $r \approx
5$~nm. The reduction of this characteristic length scale reflects
the progressive compression of the inter-cluster correlation length
under pressure.
Despite these quantitative changes, the qualitative clustering character—a peak above unity at short inter-chain separations followed by a gradual return to random mixing—remains preserved across the full pressure range.
%..............................................
\subsubsection{P--H CM-RDF: $g_{PH}^{CM}(r)$}
\label{section:PH-CM-rdf}
In sharp contrast to P--P and H--H RDFs which are depleted, 
$g_{PH}^{CM}(r)$ overshoots
the saturation value unity for short
distances. Figure~\ref{fig:cm-rdf}~(c) shows high density of
P--H  pairs at distances $r \approx 0$   reflecting genuine interpenetration of P and H chains driven by electrostatic attraction between opposite charges. The calculations of
$g_{PH}^{CM}(r)$ are validated by the trajectory snapshots showing the
center of masses coming arbitrarily close to each other, as there is
no excluded-volume restriction on center of masses. At low pressure
the larger available volume allows chains to interpenetrate freely, resulting
in $g_{PH}^{CM}(r) \approx 1.4$ for $0 < r < 2.0$~nm, approaching the
saturation value around $r \approx 4$~nm. As pressure 
increases, the random mixing occurs at shorter distances and the
interpenetration signal grows: at $P = 12.0$~bar, $g_{PH}^{CM}(r) > 1$
for short distances, reflecting the increased frequency of P--H interpenetration
under compression. For $r \gtrsim 6.0$~nm the pressure dependence becomes irrelevant,
and all curves converge to unity — the same saturation scale as
$g_{PP}^{CM}(r)$, confirming that beyond this length scale the system 
behaves as a regular mixture.

The enhanced short-range \(g_{PH}^{\rm CM}(r)\) also provides a qualitative structural signature of the strong preferential ProT\(\alpha\)--H1 association reported experimentally~\cite{borgia2018extreme,chowdhury2023driving}. We emphasize, however, that the present many-chain simulations do not provide a quantitative measure of the experimentally reported binding affinity. Determination of an equilibrium binding constant would require a separate free-energy or binding-equilibrium calculation under corresponding solution conditions.
%----------------------------------------------------------------------
\subsection{Unified structural picture}
\label{sec:unified}

Taken together, the gyration radii, the bead-level and CM-level RDFs reveal a self-consistent and
physically transparent structural picture of the binary ProT$\alpha$--Histone system
under compression.
At the bead level, the RDF signal at short r is dominated by fixed
intra-chain geometry for like-species pairs,
whereas the unlike-species RDF remains close to the random-mixing limit with only weak short-range correlations.
The pressure dependence of all bead-level peak heights is
quantitatively accounted for by a single mechanism — density normalization under
volume compression — with no evidence of structural reorganization in either species.
At the chain level, the CM-RDFs reveal the true inter-molecular organization: P--P
repulsion, H--H clustering, and P--H mixing. These qualitative characters are
mechanically robust and persist without change across the full pressure range.

The simulations do not include explicit counterions; instead, ionic screening is represented implicitly through the Debye--H\"uckel interaction. Consequently, counterion condensation, ion-mediated correlations, and counterion-release effects are not represented explicitly.
%Crucially, the absence of counter-ions in the simulation means that all electrostatic
%interactions — the repulsion between like-charged ProT$\alpha$ chains, the repulsion
%between like-charged Histone chains, and the attraction between oppositely charged
%P--H pairs — are fully unscreened.

Despite this, the system does not phase-separate and the P--H
correlations remain close to the random-mixing limit at larger
separations, underscoring that the competing effects of like-species
repulsion, unlike-species attraction, and G$\bar{o}$-native-contact mediated
intra-Histone cohesion collectively stabilize a mixed state.
The hierarchy of these
interactions, and their evolution under compression, provides a physically coherent
framework for understanding the pressure response of this binary IDP--globular protein
system.
\subsection{Diffusion in Prothymosin-Histone Condensate}
\label{sec:diffusion}
\begin{figure*}[htb]
\centering
\includegraphics[width=0.32\textwidth]{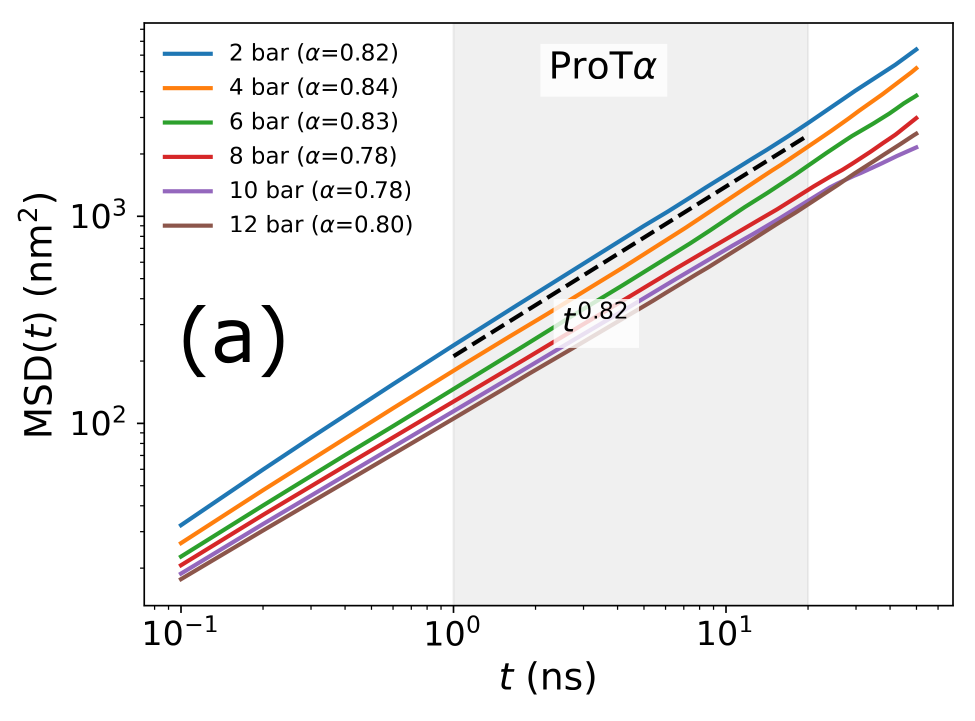}
\includegraphics[width=0.32\textwidth]{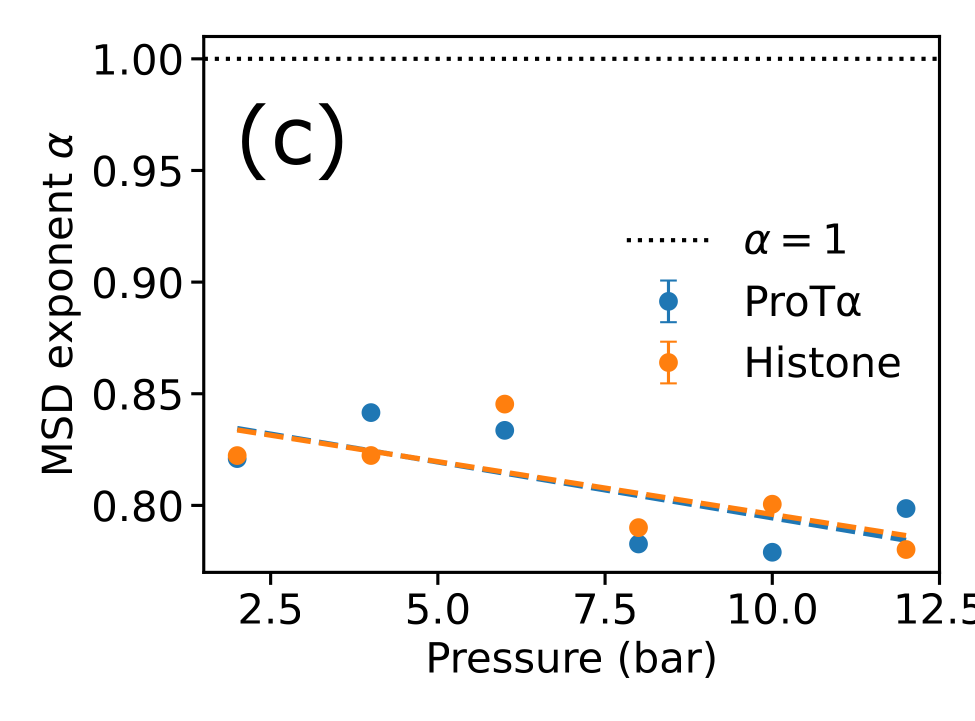}
\includegraphics[width=0.32\textwidth]{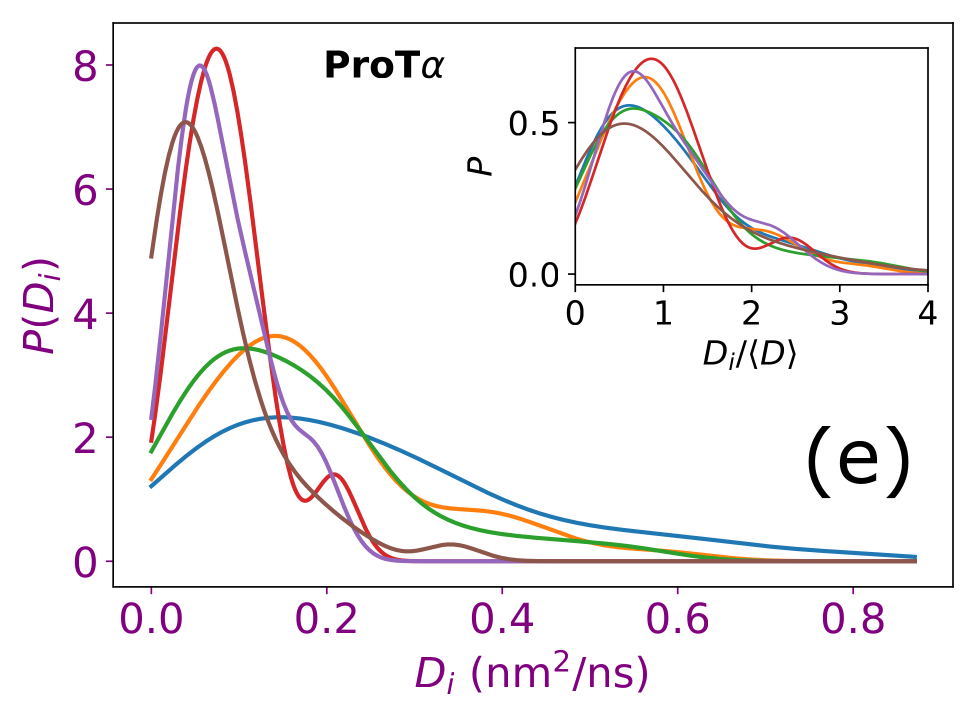}
\includegraphics[width=0.32\textwidth]{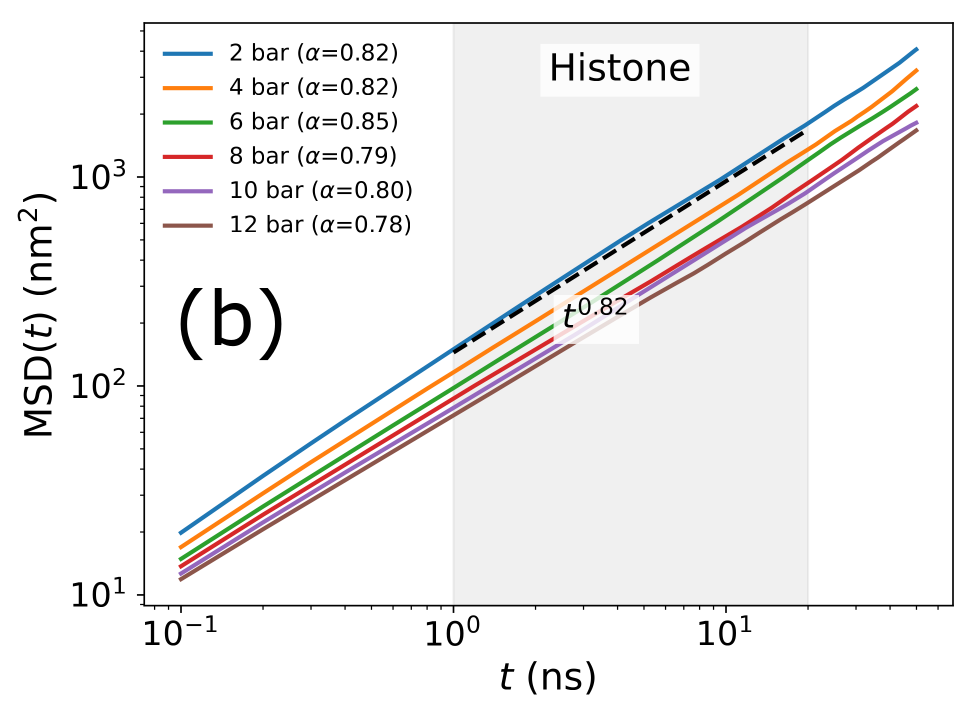}
\includegraphics[width=0.32\textwidth]{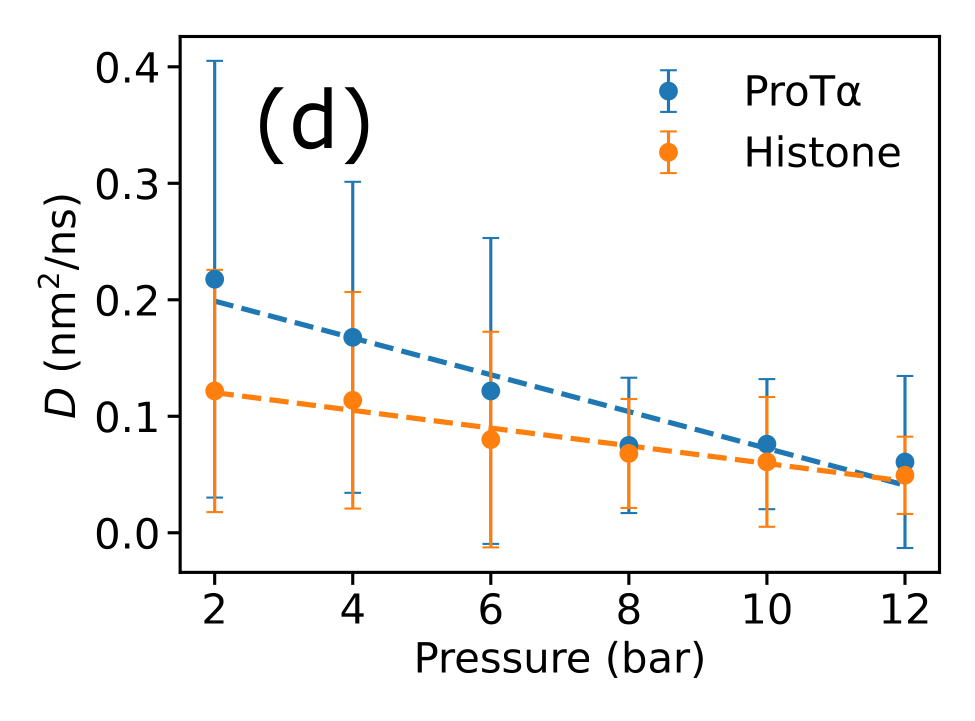}
\includegraphics[width=0.32\textwidth]{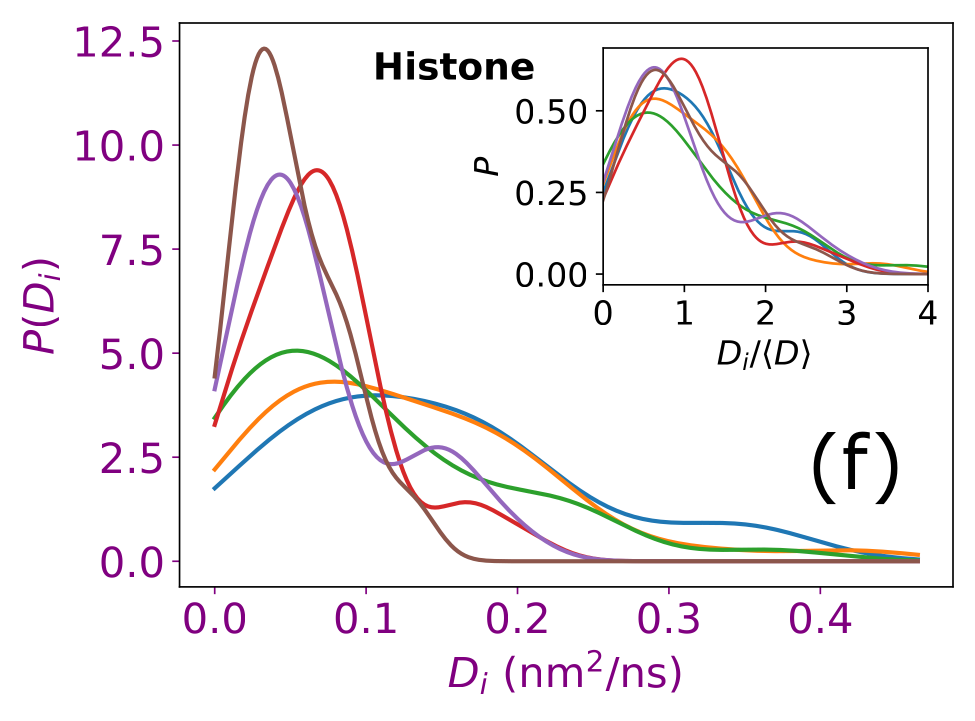}
\caption{\small
\textbf{Pressure-dependent translational dynamics of ProT$\alpha$ and Histone H1 in the condensate, showing persistent subdiffusive motion, reduced mobility under compression, and broad chain-to-chain dynamical heterogeneity.}
(a,b) Center-of-mass mean-square displacement (MSD) of ProT$\alpha$ (a) and Histone H1 (b) for pressures $P=2$--12 bar. The shaded region ($1\le t\le20$ ns) indicates the fitting window used to determine the effective diffusion coefficient $D$ and MSD exponent $\alpha$ from
$\mathrm{MSD}(t)=6Dt^\alpha$; the dashed line ($\propto t^{0.82}$) is shown as a guide to the eye.
(c,d) Pressure dependence of the the MSD exponent $\alpha$ (c) and effective diffusion coefficient $D$ (d). Error bars denote the standard deviation of the per-chain diffusion coefficients and the uncertainty of the MSD fits, respectively. Compression progressively reduces $D$, whereas $\alpha$ remains nearly constant ($\alpha\approx0.8$), indicating that pressure primarily slows molecular motion while preserving the underlying transport mechanism.
(e,f) Kernel density estimates (KDEs) of the per-chain diffusion coefficients $D_i$ for ProT$\alpha$ (e) and Histone H1 (f). Increasing pressure shifts the distributions toward lower mobilities with only modest changes in their shape. Insets show the normalized distributions, $D_i/\langle D\rangle$, which approximately collapse onto a common master curve, indicating that compression largely rescales the characteristic mobility while preserving the broad dynamical heterogeneity of the condensate.
}
\label{fig:MSD}
\end{figure*}

The diffusion of ProT$\alpha$ and 
Histone center-of-mass (CM) positions was characterized 
by computing the mean square displacement (MSD) of the CM 
trajectory for each chain. The MSD was obtained by averaging 
over all time origins (Eqn.~\ref{eq:correlation}) and all molecules of each species, 
using the CM positions reconstructed with proper periodic 
boundary condition unwrapping to ensure time-continuity 
of the displacement. As the systems did not reach fully Fickian asymptotic regime, the
effective diffusion coefficients were extracted from finite-time MSD
fits using the equation
%%%%%%%%%%%%%%%%%%%%%%%%%%%%%%%%%%%%%%%%%%%%%%%%
\begin{equation}
  MSD(t) = 6Dt^{\alpha}.
  \label{eqn:MSD}
\end{equation}
%%%%%%%%%%%%%%%%%%%%%%%%%%%%%%%%%%%%%%%%%%%%%%%%
As expected for a crowded condensate environment, both ProT$\alpha$ and Histone
exhibit subdiffusive dynamics with an anomalous diffusion
exponent $\alpha \approx 0.78$--0.82 over the entire pressure range
studied as shown in Fig.~\ref{fig:MSD}(a)-(b).
Such subdiffusive behavior is commonly observed in protein condensates and is
generally attributed to the viscoelastic nature of the dense protein
environment~\cite{watanabe2025diffusion,metzler2014anomalous}.
The weak
pressure dependence of $\alpha$ (Fig.~\ref{fig:MSD}(c)) indicates that compression alters the overall
mobility more strongly than the underlying transport mechanism. Similar values
of $\alpha$ for ProT$\alpha$ and Histone further suggest that the subdiffusion
is a collective property of the condensate rather than a
chain-specific effect.
%%%%%%%%%%%%%%%%%%%%%%%%%%%%%%%%%%%%%%%%%%%%%%%%

The pressure dependence of the diffusion coefficient and the
sub-diffusive exponent $\alpha$ is shown in
Fig.~\ref{fig:MSD}(c)-(d).
The diffusion coefficients decrease 
monotonically with increasing pressure for both P and H, 
ranging from $D_P \approx 0.22$~nm$^2$/ns 
at $P = 2$~bar to $D_P \approx 0.06$~nm$^2$/ns 
at $P = 12$~bar, and $D_H \approx 0.12 $ to 
$0.05 $~nm$^2$/ns over the same range. The 
decrease reflects increased molecular crowding as the box 
volume is compressed, which slows the translational motion 
of both species. Notably, the two curves converge at high 
pressure, suggesting that the dynamical distinction between 
P and H is suppressed under strong confinement.
ProT$\alpha$ consistently diffuses faster than 
Histone at all pressures ($D_P > D_H$), consistent with 
P having a smaller radius of gyration ($\langle R_g \rangle_P 
\approx 4.3$~nm versus $\langle R_g \rangle_H \approx 5.2$~nm) 
and the presence of a folded globular domain absent in ProT$\alpha$. The ratio $D_P/D_H \approx 1.5$--$2$ at low pressure 
decreasing toward unity at high pressure further supports 
a pressure-driven equalization of the two species' mobilities.

A prominent feature of Fig.~\ref{fig:MSD}(d) is the large
standard deviation for both species at all pressures. This molecule-to-molecule variability
reflects the heterogeneous dynamics characteristic of dense viscoelastic
condensates~\cite{FUS_viscoelastic}.
Individual chains experience different local environments and
interaction histories, leading to substantial variations in their
effective mobilities. We emphasize that this breadth is
\emph{independent} of the subdiffusive regime: even in a Fickian
system, dynamical heterogeneity from chain to chain would produce a
broad $P(D_i)$. The quantity $D_i$ extracted here is therefore an effective mobility
in the subdiffusive regime, and its broad distribution shown in (Figs.~\ref{fig:MSD}(e)-(f)) reflect genuine chain-to-chain heterogeneity arising from fluctuations in local density, transient intermolecular contacts, and the enhanced short-range H--H correlations revealed by the CM-RDF (Sec.~\ref{section:HH-CM-rdf}).
%%%%%%%%%%%%%%%%%%%%%%%%%%%%%%%%%%%%%%%%%
\begin{figure}[htb]
\includegraphics[width=0.98\linewidth]{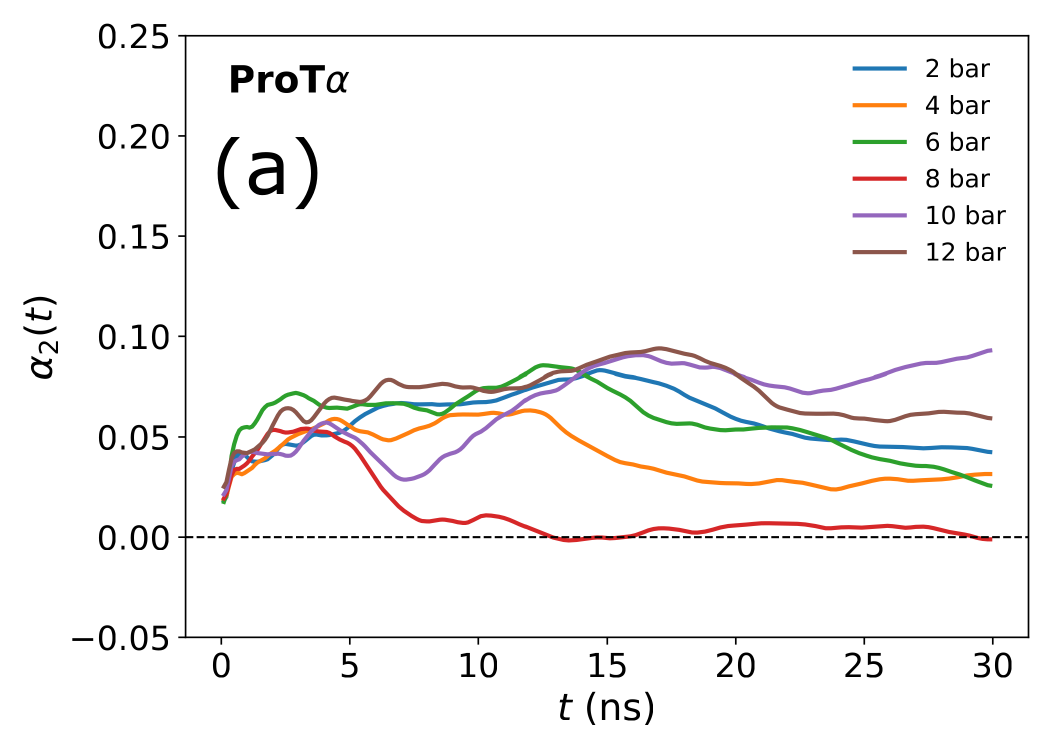}
\includegraphics[width=0.98\linewidth]{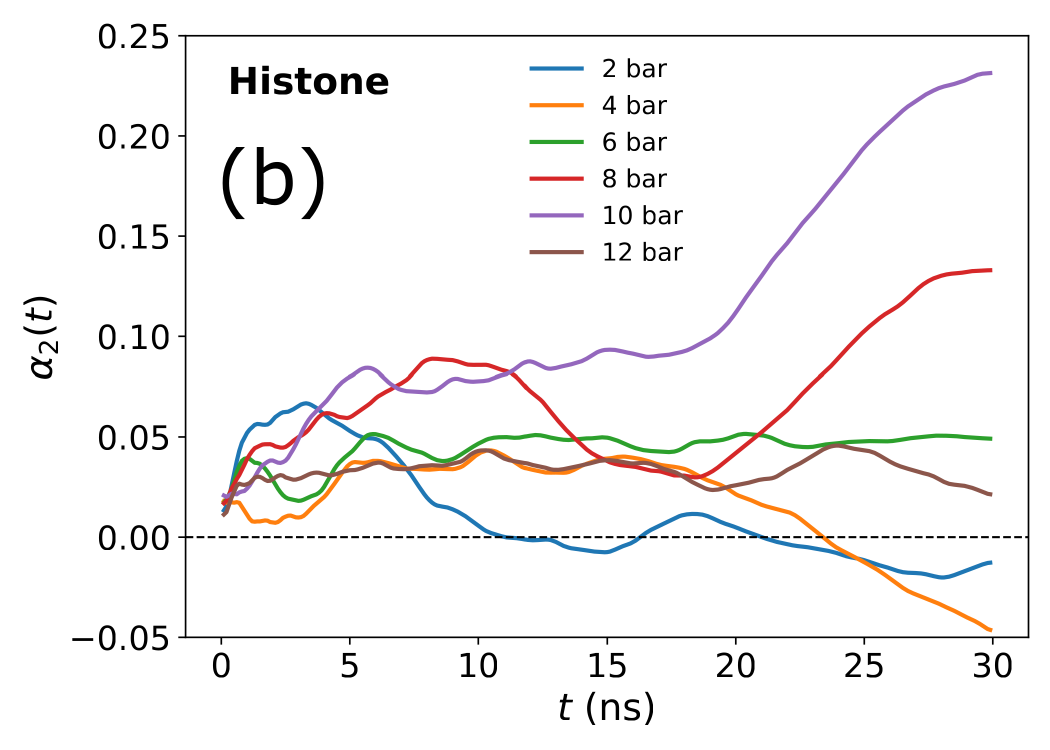}
\caption{\small Non-Gaussian parameter $\alpha_2(t) = 3\langle r^4(t)\rangle /
5\langle r^2(t)\rangle^2 - 1$ for ProT$\alpha$ (top) and Histone H1
(bottom) at pressures $P = 2$--$12$~bar. A positive $\alpha_2(t)$
indicates heterogeneous, non-Gaussian displacement statistics
characteristic of a viscoelastic environment. The effect is
substantially stronger and more persistent for Histone than for
ProT$\alpha$, consistent with the broader distribution of local
environments sampled by Histone chains within the condensate.}
\label{fig:Non-Gaussian}
\end{figure}
%%%%%%%%%%%%%%%%%%%%%%%%%%%%%%%%%%%%%%%%%%%%%%%%
The diffusion coefficients obtained here,
$D \approx 0.05$--$0.22$~nm$^2$/ns, should be regarded as
effective mobilities within the present coarse-grained dynamical
model rather than as quantitatively mapped experimental diffusion
coefficients. For comparison, atomistic simulations of FUS
low-complexity domain (FUS-LCD) condensates reported diffusion
coefficients of $D \approx 0.26\times10^{-3}$~nm$^2$/ns, comparable
to experimental NMR diffusometry measurements
\cite{FUS_sim,Fawzi}. The substantially larger absolute values
obtained in the present coarse-grained simulations are consistent
with the accelerated dynamics expected from coarse graining and
the absence of explicit solvent-mediated hydrodynamic interactions,
but we do not assign a quantitative conversion factor between the
two descriptions. The more robust result is the systematic decrease
of mobility with increasing packing fraction for both protein
species.

Quantitative differences between the present ProT$\alpha$--Histone
system and FUS-LCD condensates are nevertheless expected. Unlike the
weakly charged FUS-LCD chains (net charge $\approx -2$), ProT$\alpha$
and Histone H1 carry large opposite charges ($-44$ and $+53$,
respectively), resulting in strong electrostatic coupling and
continuous transient intermolecular associations. Despite these
differences, both systems exhibit the same qualitative picture of slow,
subdiffusive transport within a dense condensate environment.

From the diffusion coefficients for the individual chains $D_i$s, we have calculated
the Kernel Density Estimates (KDEs) for both ProT$\alpha$
and Histone proteins show in Fig.~\ref{fig:MSD}(e)-(f) of the per-chain
diffusion coefficient distributions $P(D_i$s at different pressures. The main panel show
the distributions on the absolute diffusion scale, $D_i$, revealing a
pressure-induced shift toward lower mobility. Insets show the same data
scaled by the species- and pressure-dependent mean diffusion coefficient,
$D_i/\langle D\rangle$. The approximate collapse of the scaled distributions
suggests that pressure primarily rescales the mobility
rather than qualitatively altering the underlying
distribution of chain-to-chain diffusivities.
In addition, we have also looked at the non-Gaussian parameter
$\alpha_2(t)$ (Eqn.~\ref{Eqn:Non-Gaussian}) shown in
Fig.~\ref{fig:Non-Gaussian}.

\begin{equation}
\alpha_2(t)=\frac{3\langle r^4(t)\rangle}
{5\langle r^2(t)\rangle^2}-1
\label{Eqn:Non-Gaussian}
\end{equation}

The non-Gaussian displacement statistics implied by this broad
distribution are consistent with recent simulations of FUS-LCD and
DDX4 droplets that report anomalous subdiffusion at intermediate
time scales and spatially heterogeneous diffusivity between the
droplet interior and interface~\cite{watanabe2025diffusion}.

$\alpha_2(t)$ extracted from our simulations
(Fig.~\ref{fig:Non-Gaussian}) remains positive over the
intermediate-time window, indicating heterogeneous displacement
statistics. The effect is substantially stronger for Histone than
for ProT$\alpha$, consistent with the broader distribution of local
environments sampled by Histone chains within the condensate.
Because $\alpha_2(t)$ depends on the fourth moment of the
displacement distribution, $\langle r^4(t)\rangle$, it is
particularly sensitive to finite-sampling effects. The weak
oscillatory structure in the curves is therefore likely a
consequence of limited statistics.
\subsection{Chain relaxation dynamics}
\label{sec:relax}
\begin{figure}[ht]
\centering
\includegraphics[width=0.48\textwidth]{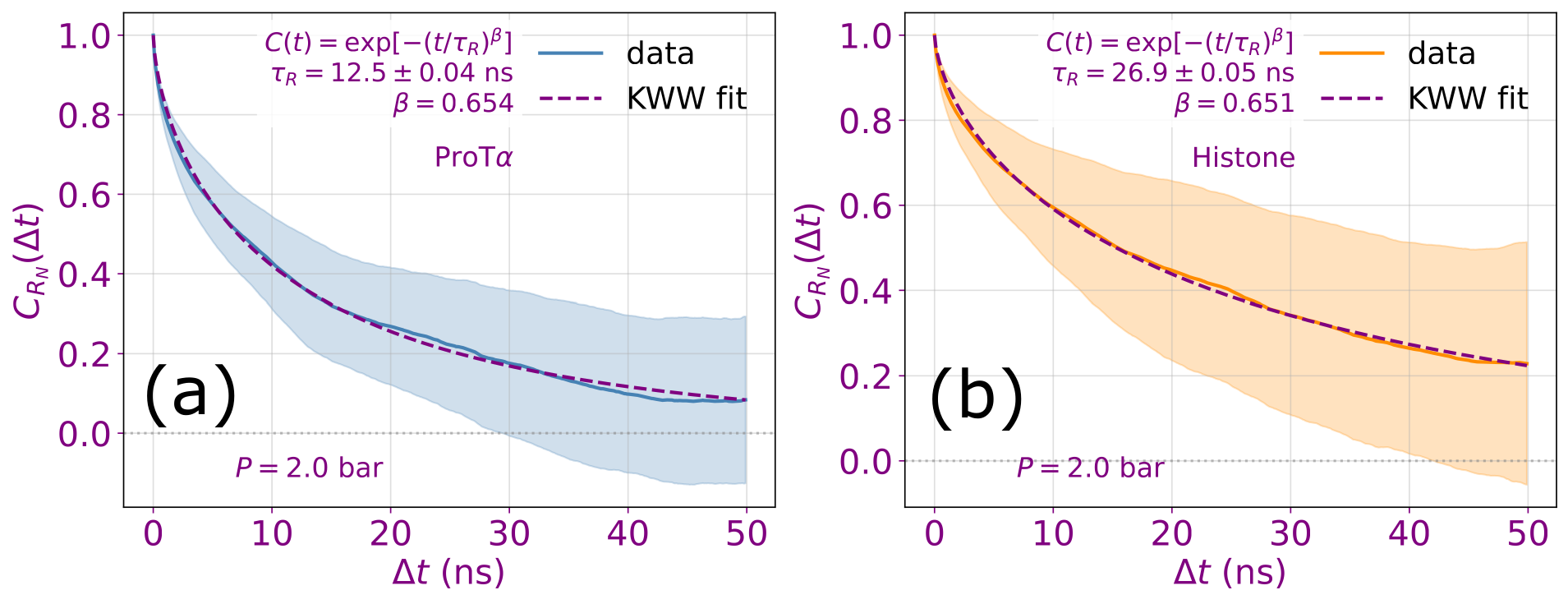}
\includegraphics[width=0.48\textwidth]{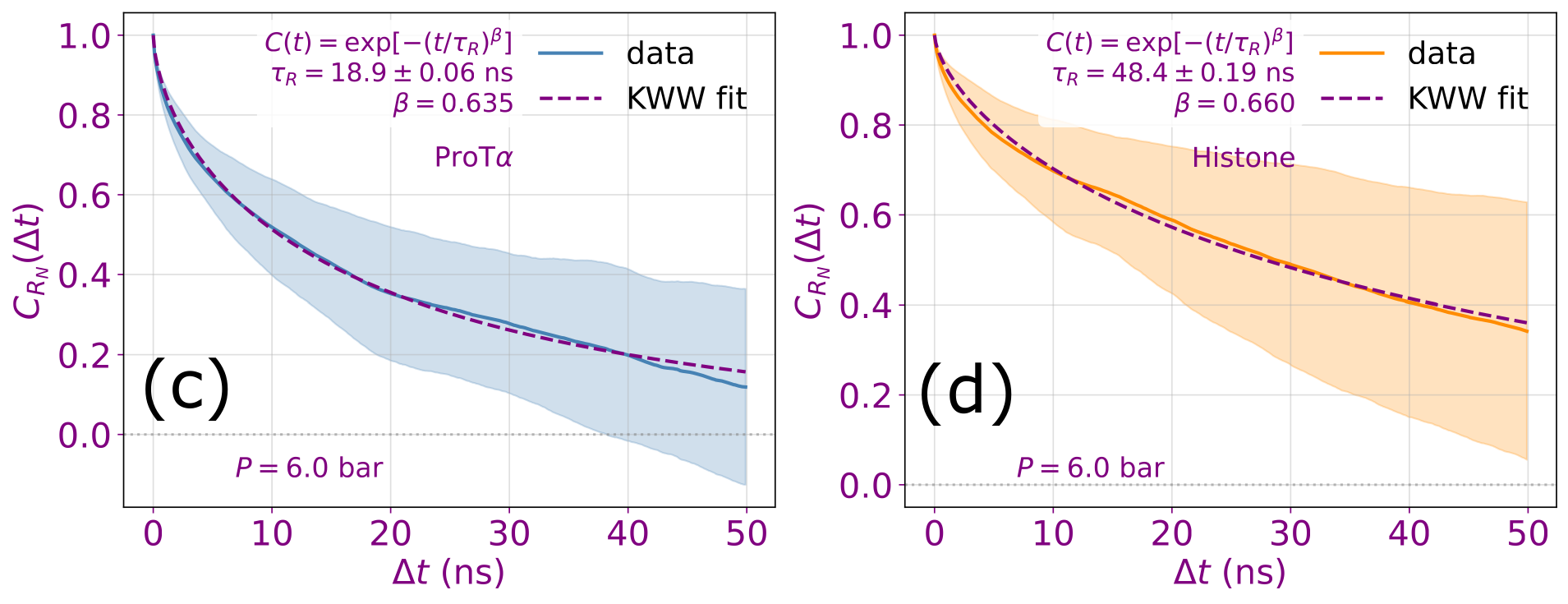}
\includegraphics[width=0.48\textwidth]{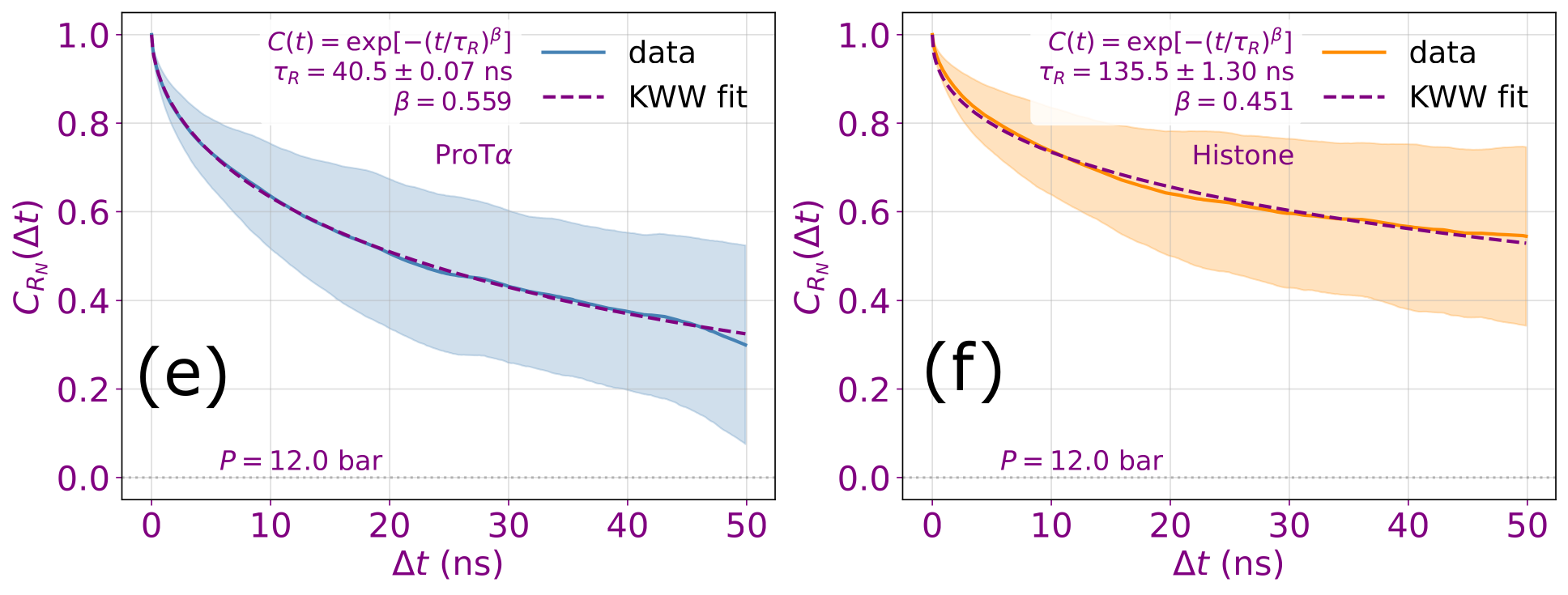}
\includegraphics[width=0.23\textwidth]{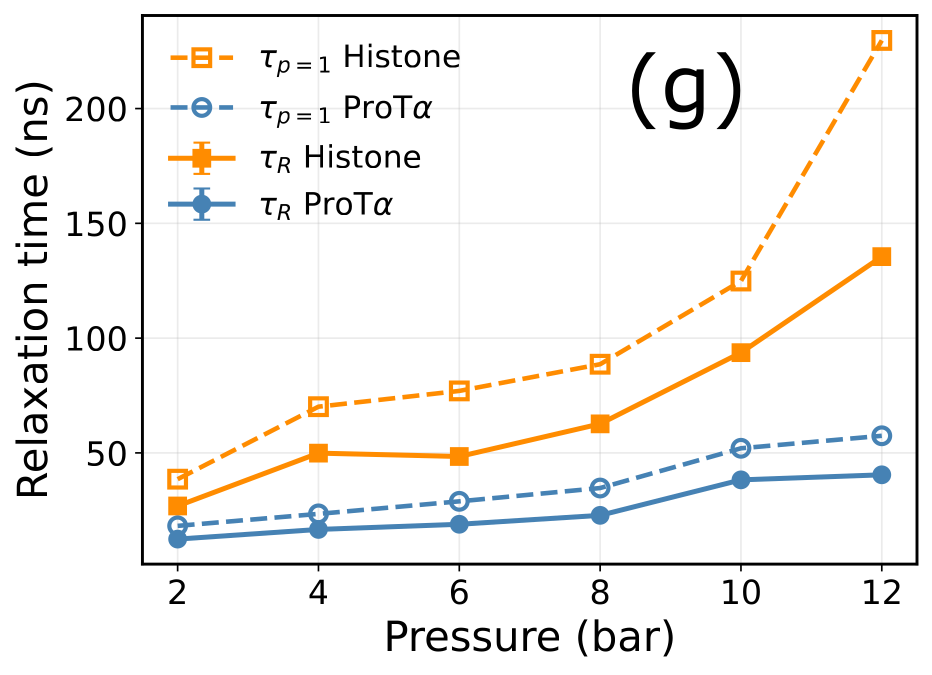}
\includegraphics[width=0.23\textwidth]{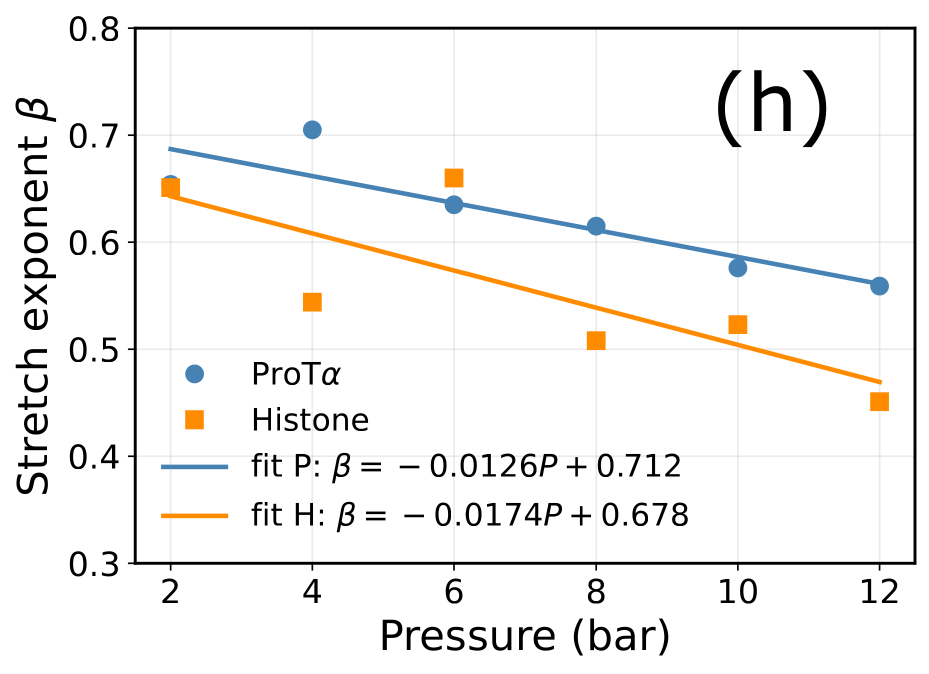}
\caption{\small End-to-end vector autocorrelation $C(t) =
\langle\mathbf{R}_N(t)\cdot\mathbf{R}_N(0)\rangle/\langle R_N^2\rangle$
for ProT$\alpha$ (a,c,e) and Histone (b,d,f) at $P = 2.0$, $6.0$,
and $12.0$~bar. Solid lines are simulation data with shaded
bands indicating chain-to-chain standard deviation; dashed lines
are KWW stretched-exponential fits
$C(t) = \exp[-(t/\tau_R)^\beta]$ with fit parameters shown.
Both species show clearly non-exponential decay ($\beta < 1$)
with relaxation times and stretch exponents that depend
systematically on pressure (Table~\ref{tab:rouse_modes}).
Pressure dependence of chain relaxation dynamics. (g,h)
 relaxation time $\tau_R$ and slowest Rouse mode $\tau_{p=1}$ (g),
  and stretched exponential coefficient $\beta$ (h). All quantities shown for P
  (ProT$\alpha$, blue circles and squares) and H (Histone, orange
  circles and squares) as a function of pressure.}
\label{fig:rn-relax}
\end{figure}

We now analyze different time scales of relaxation of the individual chains
and the subchain dynamics from the simulation data and using Rouse
model~\cite{rubinstein2003polymer}. 
\paragraph{End-to-end vector correlation $C(t)$:}~We start with the 
end-to-end vector autocorrelation function
\begin{equation}
C(t) = \langle \mathbf{R}_N(t)\cdot\mathbf{R}_N(0)\rangle / 
\langle R_N^2\rangle
\label{eqn:Tau-Rn}
\end{equation}
shown in
Fig.~\ref{fig:rn-relax} that captures the relaxation of the entire chain and is 
directly comparable to reconfiguration times measured by 
single-molecule FRET~\cite{Schuler_smFRET} both for 
%%%%%%%%%%%%%%%%%%%%%%%%%%%%%%%%%%%%%%%%%%%%%
%%%%%%%%%%%%%%%%%%%%%%%%%%%%%%%%%%%%%%%%%%%%
Prot$\alpha$ and Histone at three 
pressures. Both species show a clearly non-exponential 
decay, well described by the 
Kohlrausch-Williams-Watts (KWW) stretched exponential 
$C(t) = \exp[-(t/\tau_R)^\beta]$ with $\beta \approx 0.6 < 1$, 
reminiscent of heterogeneous viscoelastic dynamics.
The relaxation times increase strongly with pressure
(Fig.~\ref{fig:rn-relax}(g)); $\tau_R^P$ grows from $12.5$ ns at $P=2$
bar to $40.5$ ns at $P=12$ bar, while $\tau_R^H$ grows from $26.9$ ns
to $135.5$ ns over the same range. Histone relaxes consistently slower
than ProT$\alpha$ at all pressures ($\tau_R^H/\tau_R^P \approx
2$--$3.5$), reflecting both its longer chain length and the more
constrained local environment associated with the enhanced H--H
correlations observed in the CM-RDF. The stretch exponent $\beta$
shows (Fig.~\ref{fig:rn-relax}(h)) an overall decreasing trend with pressure, from approximately
$0.65$ at $P=2$ bar to $0.45$--$0.56$ at $P=12$ bar
(Table~\ref{tab:rouse_modes}), with moderate non-monotonicity at
intermediate pressures within the fitting uncertainty. Histone
generally exhibits smaller $\beta$ values than ProT$\alpha$,
indicating a broader distribution of relaxation times and more
heterogeneous relaxation dynamics.
\iffalse
%%%%%%%%%%%%%%%%%%%%%%%%%%%%%%%%%%%%%%%%%%%%%%%%%%%%%%%%
\begin{figure}[ht]
\centering
\includegraphics[width=0.48\textwidth]{PDF/Relax/relax_vs_pressure.pdf}
\caption{\small Pressure dependence of chain relaxation dynamics.
  (top): relaxation time $\tau_R$ and slowest Rouse mode $\tau_{p=1}$,
  and (bottom): stretched exponential coefficient $\beta$. All quantities shown for P
  (ProT$\alpha$, blue circles and squares) and H (Histone, orange
  circles and squares) as a function of pressure.}
\label{fig:relax_vs_P}
\end{figure}
%%%%%%%%%%%%%%%%%%%%%%%%%%%%%%%%%%%%%%%%%%%%%%%%%%%%
\fi
\begin{table}[ht]
\footnotesize 
\centering 
\caption{Pressure dependence of chain relaxation parameters 
for ProT$\alpha$ (P) and Histone (H). 
$\tau_R$ is from $C(t) = \exp[-(t/\tau_R)^\beta]$, 
$\beta$ is the stretch exponent, $\tau_{p=1}$ is the 
slowest Rouse mode. Uncertainties in $\tau_R$ from fit covariance.}
\label{tab:rouse_modes}
\begin{tabular}{ccccccc}
\hline\hline 
$P$ & $\tau_R^P$ & $\beta_P$ & $\tau_R^H$ & $\beta_H$ 
    & $\tau_{p=1}^P$ & $\tau_{p=1}^H$ \\
(bar) & (ns) & & (ns) & & (ns) & (ns) \\
\hline 
 2 & $12.5\pm0.04$ & 0.654 & $ 26.9\pm0.05$ & 0.651 &  18.2 &  38.6 \\
 4 & $16.7\pm0.07$ & 0.705 & $ 49.9\pm0.25$ & 0.544 &  23.4 &  70.1 \\
 6 & $18.9\pm0.06$ & 0.635 & $ 48.4\pm0.19$ & 0.660 &  28.9 &  77.0 \\
 8 & $22.8\pm0.05$ & 0.615 & $ 62.6\pm0.14$ & 0.508 &  34.7 &  88.6 \\
10 & $38.3\pm0.11$ & 0.576 & $ 93.7\pm0.27$ & 0.523 &  52.0 & 124.9 \\
12 & $40.5\pm0.07$ & 0.559 & $135.5\pm1.30$ & 0.451 &  57.5 & 229.7 \\
\hline\hline 
\end{tabular}
\end{table}
%%%%%%%%%%%%%%%%%%%%%%%%%%%%%%%%%%%%%%%%%%%%%%%%%%%%%%
%%%%%%%%%%%%%%%%%%%%%%%%%%%%%%%%%%%%%%%%%%%%%%%%%%%%%%%%
\paragraph{Rouse mode Spectrum:}
In the Rouse model, chain dynamics are decomposed into
normal modes with relaxation times $\tau_p \sim p^{-2}$,
where $p = 1$ is the longest internal mode and $p > 1$
probes progressively shorter length scales~\cite{rubinstein2003polymer}.
For a Gaussian chain $\tau_1$ closely tracks
the end-to-end vector relaxation $\tau_R$~\cite{rubinstein2003polymer},
and any departure between them signals dynamics beyond
simple Rouse behavior.
%%%%%%%%%%%%%%%%%%%%%%%%%%%%%%%%%%%%%%%%%%%%%%%%%%%%%%%%
\begin{figure}[ht]
\centering
\includegraphics[width=0.48\textwidth]{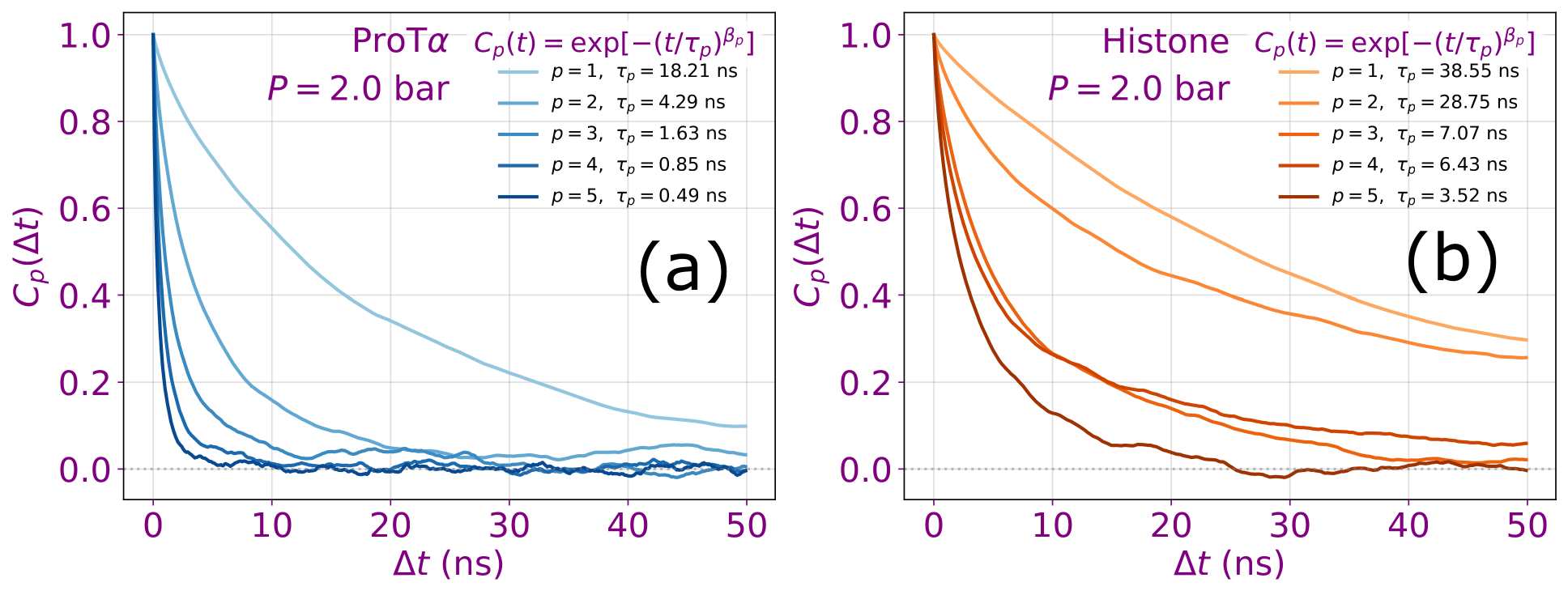}
\includegraphics[width=0.48\textwidth]{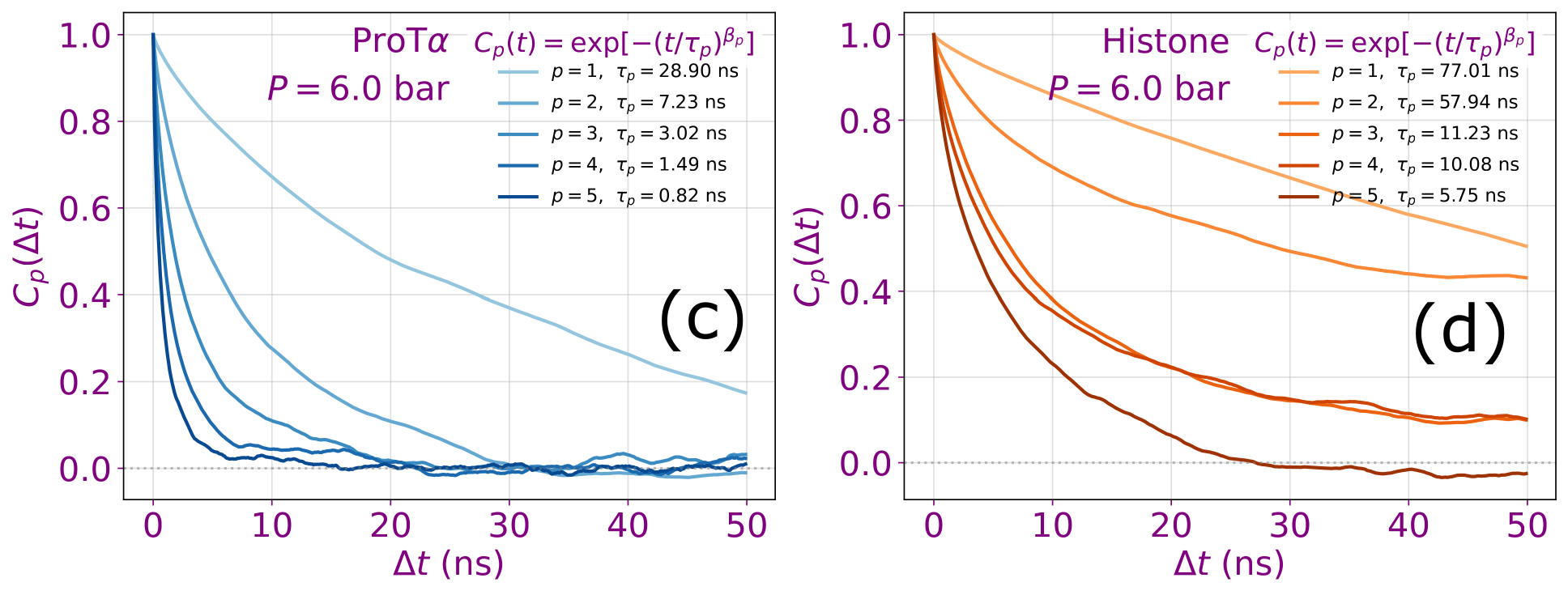}
\includegraphics[width=0.48\textwidth]{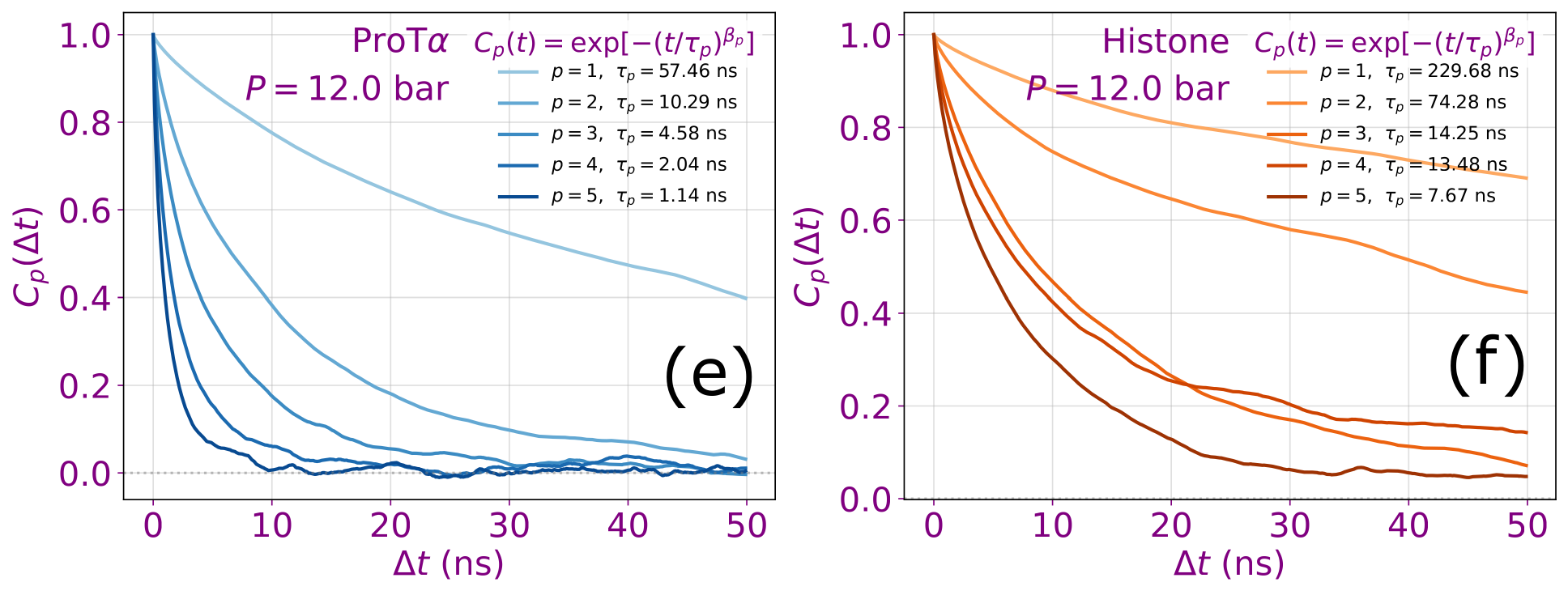}
\includegraphics[width=0.23\textwidth]{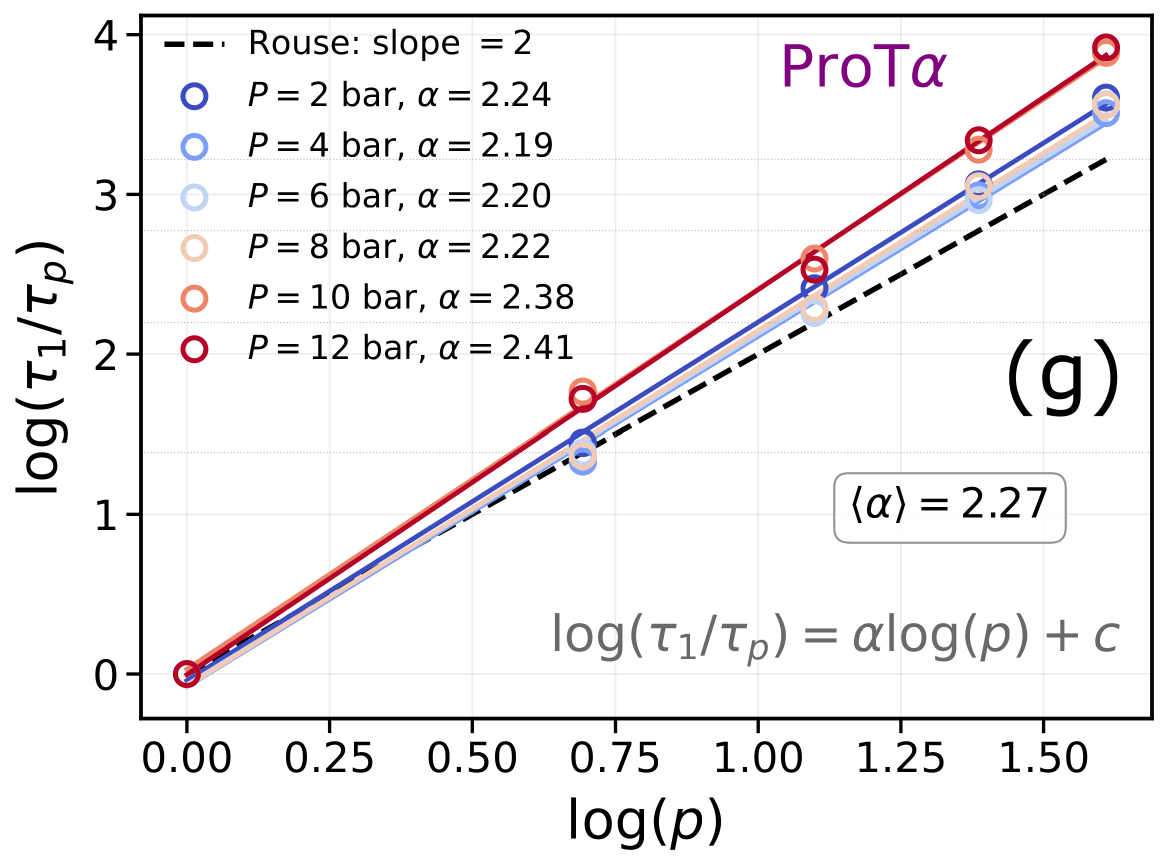}
\includegraphics[width=0.23\textwidth]{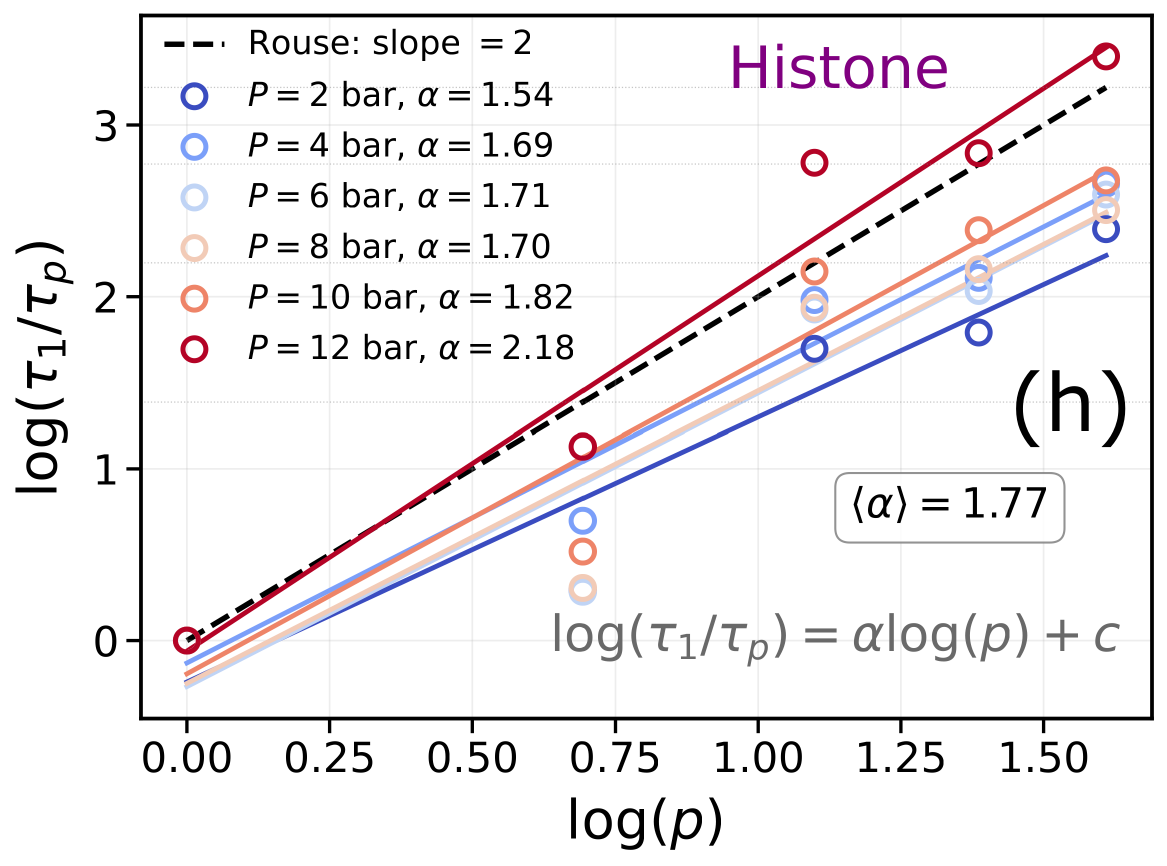}
\caption{\small Relaxation functions $C_p(t)$ of the first five Rouse modes
($p=1$--5) for ProT$\alpha$ (left) and Histone H1 (right) at
representative pressures. Solid lines are stretched-exponential fits
used to extract the mode relaxation times reported in
Table~\ref{tab:rouse_modes}. Relaxation becomes progressively slower
with increasing pressure for both proteins. Histone exhibits
substantially slower relaxation than ProT$\alpha$, particularly for the
lowest modes, with the $p=1$ and $p=2$ modes remaining comparable to
the overall chain relaxation time $\tau_R$.
(last row) Rouse mode scaling for ProT$\alpha$
(top) and Histone (bottom) at P=2--12 bar.
Log-log plot of the normalized relaxation time $\tau_1/\tau_p$ 
 versus mode index $p$, where normalization by $\tau_1$ collapses all curves to a common
origin. Solid lines are power-law fits
$\log\left(\tau_1/\tau_p\right) = \alpha\log(p) + c$;
the dashed line is the Rouse prediction (slope =2). 
Colors progress from blue (P=2~bar) to red (P=12~bar).}
\label{fig:rouse_modes}
\end{figure}
%%%%%%%%%%%%%%%%%%%%%%%%%%%%%%%%%%%%%%%%%%%%%%%%%%%%%%%%
\iffalse
%%%%%%%%%%%%%%%%%%%%%%%%%%%%%%%%%%%%%%%%%%%%%%%%%%%%%%%%
\begin{figure}[ht]
\centering
\includegraphics[width=0.48\textwidth]{PDF/Relax/rouse_mode_scaling.pdf}
\caption{\small Rouse mode scaling for ProT$\alpha$
(top) and Histone (bottom) at P=2--12 bar.
Log-log plot of the normalized relaxation time $\tau_1/\tau_p$ 
 versus mode index $p$, where normalization by $\tau_1$ collapses all curves to a common
origin. Solid lines are power-law fits
$\log\left(\tau_1/\tau_p\right) = \alpha\log(p) + c$;
the dashed line is the Rouse prediction (slope =2). 
Colors progress from blue (P=2~bar) to red (P=12~bar).}
\label{fig:rouse_scaling}
\end{figure}
\fi
%%%%%%%%%%%%%%%%%%%%%%%%%%%%%%%%%%%%%%%%%%%%%%%%%%%%%%%%

The relaxation times of the individual Rouse modes are summarized in
Table~\ref{tab:rouse_modes} and illustrated in
Figs.~\ref{fig:rouse_modes}(a)-(h). For both
proteins the relaxation times increase systematically with pressure,
consistent with the slowing of the overall chain dynamics discussed
above. ProT$\alpha$ exhibits the expected hierarchy of relaxation
times, with progressively faster relaxation for increasing mode number.
Histone displays a similar overall trend, but the separation between
the lowest modes is noticeably reduced. In particular, the $p=1$ and
$p=2$ modes remain comparable to the end-to-end relaxation time
$\tau_R$ over the entire pressure range, indicating a stronger coupling
between global chain relaxation and the slowest internal modes than in
ProT$\alpha$.

Fig.~\ref{fig:rouse_modes}(g) further shows that the mode spectrum of
ProT$\alpha$ is broadly consistent with a Rouse-like ordering of
relaxation times, whereas Histone exhibits a weaker dependence on mode
number for the lowest modes (Fig.~\ref{fig:rouse_modes}(h)). This deviation is most apparent for
$p=1$--2 and becomes increasingly pronounced with pressure. The effect
is consistent with the composite architecture of Histone H1, consisting
of a folded globular domain connected to disordered charged tails, as
well as the enhanced short-range H--H correlations revealed by the
CM-RDF. Nevertheless, higher modes relax substantially faster than the
global chain relaxation, indicating that local conformational
rearrangements remain dynamic even under the highest compression
studied.
\subsection{Contact lifetimes and the fast-exchange regime}
\label{sec:contacts}
To characterize the timescale of specific inter-chain interactions
we compute contact lifetimes for all three pair types — P--P,
H--H, and P--H — using a center-of-mass cutoff $r_c = 2.0$~nm,
which captures chain pairs whose centers of mass lie within
approximately half a chain radius of one another.
Table~\ref{tab:contacts} summarizes the mean contact lifetime
$\langle\tau_\text{bind}\rangle$, its standard error,
and the ratio $n_\text{obs}/n_\text{rand}$ of observed contacts
to the random-mixing expectation
$n_\text{rand} = (N-1)(4\pi r_c^3/3)/V$
at each pressure.
\begin{table}[htbp]
  \footnotesize 
  \centering
\caption{Contact statistics for P--P, H--H, and P--H chain pairs as a
function of pressure. $\tau_\text{bind}$ is the characteristic contact
lifetime obtained from exponential fits to the survival probability
(not shown here), and $\sigma_{\tau_\text{bind}}$ is the corresponding standard error.
The ratio $n_\text{obs}/n_\text{rand}$ compares the observed number of
contacts per chain with the random-mixing expectation
$n_\text{rand}=(N-1)(4\pi r_c^3/3)/V$ at cutoff
$r_c=2.0$~nm. All lifetimes are reported in ns.}
\label{tab:contacts}
\renewcommand{\arraystretch}{1.2}
\begin{tabular}{c | cc c | cc c | cc c}
\hline\hline
 & \multicolumn{3}{c|}{P--P}
 & \multicolumn{3}{c|}{H--H}
 & \multicolumn{3}{c}{P--H} \\
$P$
 & $\langle\tau_\text{bind}\rangle$ &  $\sigma_{\tau_\text{bind}}$  & ratio
 & $\langle\tau_\text{bind}\rangle$ & $\sigma_{\tau_\text{bind}}$ & ratio
 & $\langle\tau_\text{bind}\rangle$ & $\sigma_{\tau_\text{bind}}$ & ratio \\
\hline
 2 & 0.38 & 0.04 & 0.58
   & 0.50 & 0.06 & 0.72
   & 0.43 & 0.02 & 1.31 \\
 4 & 0.42 & 0.03 & 0.48
   & 0.49 & 0.03 & 0.62
   & 0.48 & 0.02 & 1.18 \\
 6 & 0.45 & 0.03 & 0.65
   & 0.65 & 0.07 & 0.70
   & 0.49 & 0.02 & 0.97 \\
 8 & 0.44 & 0.03 & 0.57
   & 0.50 & 0.04 & 0.53
   & 0.52 & 0.02 & 1.00 \\
10 & 0.48 & 0.03 & 0.61
   & 0.58 & 0.04 & 0.76
   & 0.54 & 0.02 & 1.04 \\
12 & 0.48 & 0.03 & 0.65
   & 0.67 & 0.07 & 1.30
   & 0.56 & 0.02 & 1.01 \\
\hline\hline
\end{tabular}
\end{table}

%%%%%%%%%%%%%%%%%%%%%%%%%%%%%%%%%%%%%%%%%%%%%%%%%%%%%%%%
\begin{figure}[ht]
\centering
\includegraphics[width=0.48\textwidth]{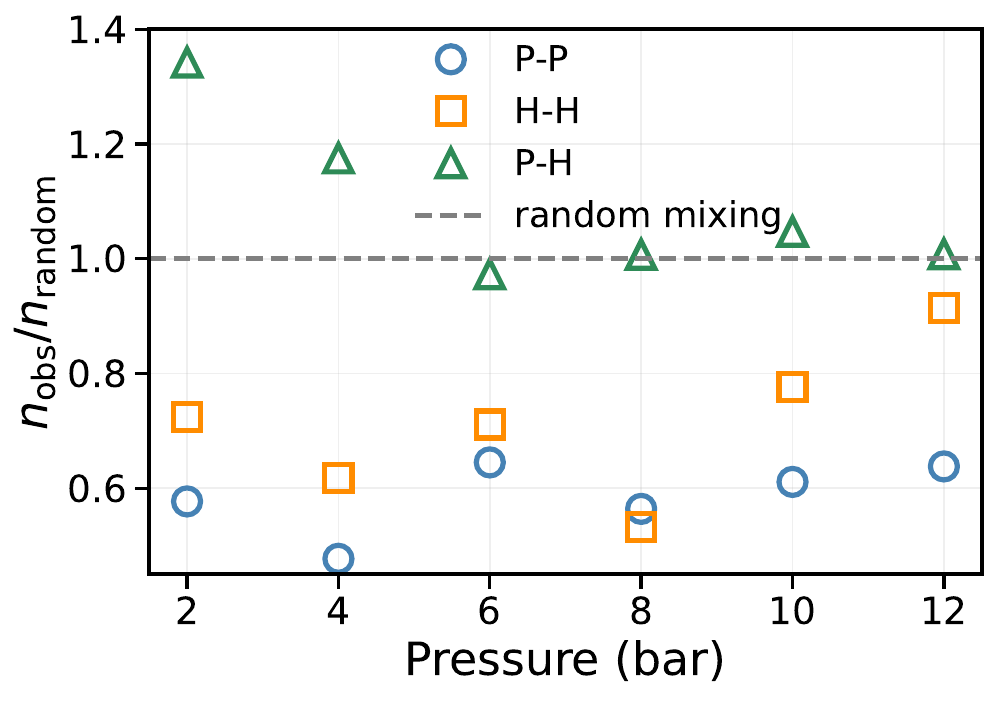}
\caption{Ratio of observed inter-chain contacts to random-mixing
expectation, $n_\text{obs}/n_\text{rand}$, for P--P, H--H, and
P--H pairs as a function of pressure at cutoff $r_c = 2.0$~nm.
P--P contacts are depleted at all pressures (electrostatic
repulsion); P--H contacts show excess at low pressure driven
by charge complementarity, converging to random mixing above
$P \approx 6$~bar; H--H values scatter around or below unity,
reflecting the balance between like-charge repulsion and
native-contact-mediated attraction.   The dashed line corresponds to the
random-mixing expectation
$n_{obs}/n_{rand}=1$.}
\vspace{-4.0pt}
\label{fig:contacts}
\end{figure}
%%%%%%%%%%%%%%%%%%%%%%%%%%%%%%%%%%%%%%%%%%%%%%%%%%%%%%%%
%%%%%%%%%%%%%%%%%%%%%%%%%%%%%%%%%%%%%%%%%%%%%%%%

The most striking result is the extremely short contact lifetime for
all pair types and pressures studied. The survival probabilities decay
rapidly on a sub-nanosecond timescale, yielding characteristic
lifetimes $\tau_\text{bind}\approx0.4$--$0.7$~ns. These values are
remarkably similar for P--P, H--H, and P--H contacts and exhibit only a
weak pressure dependence, despite the distinct structural correlations
revealed by the RDF analysis.

Since the end-to-end relaxation times satisfy
$\tau_R^P \approx 12$--$40$~ns and
$\tau_R^H \approx 27$--$136$~ns
(Table~\ref{tab:rouse_modes}), the ratio

\begin{equation}
\frac{\tau_\text{bind}}{\tau_R}
\approx 0.01\text{--}0.04
\end{equation}

demonstrates a pronounced separation of timescales between local
contact dynamics and whole-chain relaxation. Individual intermolecular
contacts therefore form and dissociate many times during a single chain
relaxation event, placing the condensate firmly in the fast-exchange
regime throughout the pressure range studied.

The condition $\tau_\text{bind}\ll\tau_R$ implies that local
intermolecular contacts act as rapidly exchanging associations rather
than persistent cross-links. Consequently, the structural organization
of the condensate emerges from collective many-chain correlations and
continuous contact rearrangement. This interpretation is consistent
with the CM-RDFs, which reveal persistent whole-chain correlations,
and with the subdiffusive dynamics discussed in
Sec.~\ref{sec:diffusion}, which arise from the viscoelastic many-body
environment rather than from long-lived intermolecular bonds.
\section{Summary and Conclusions}
\label{sec:summary}

We have characterized the structure and dynamics of a binary
ProT$\alpha$--Histone condensate under compression using
coarse-grained molecular dynamics simulations. The results
reveal a viscoelastic fluid in the fast-exchange regime,
whose properties are governed by sequence-dependent chain
architecture, the spatial distribution of charged residues,
and collective many-body crowding.

\paragraph{Structure.}
The condensate is structurally well-mixed and mechanically
robust across the full pressure range. Neither chain
compacts under compression — $\langle R_g\rangle$ is
pressure-independent for both species — and the qualitative
character of all three CM-RDFs (P--P depletion, H--H
association, P--H mixing) is preserved. The uniform
suppression of bead-level RDF peaks by the volume
compression ratio confirms that local chain geometry
is unaffected by pressure. Despite strong
electrostatic interactions, the
system does not phase-separate: the charge complementarity
between ProT$\alpha$ and Histone drives transient
contacts but not macroscopic segregation.

\paragraph{Dynamics.}
All dynamical timescales increase with pressure, but
their relative ordering and physical interpretation
are pressure-invariant. Both species show subdiffusive
MSD ($\alpha \approx 0.82$), broad single-chain diffusivity distributions
($\sigma_D \sim \langle D\rangle$), and stretched-exponential
end-to-end relaxation ($\beta_\text{KWW} \approx
0.45$--$0.65 < 1$) — all signatures of a heterogeneous
viscoelastic fluid. The Rouse mode spectrum cleanly
distinguishes the two chain architectures: ProT$\alpha$
follows near-Rouse scaling ($\langle\alpha\rangle = 2.27$)
consistent with its uniformly disordered, charge-distributed
backbone, while Histone shows sub-Rouse behavior
($\langle\alpha\rangle = 1.77$) with anomalously slow
low-order modes, consistent with the presence of the folded
globular domain and the composite globular-domain/disordered-tail
architecture of Histone H1.

\paragraph{Contact lifetimes and fast exchange.}
All inter-chain contacts—P--P, H--H, and P--H—are highly
transient, with characteristic lifetimes $\tau_\text{bind}
\approx 0.4$--$0.7$~ns, approximately two orders of magnitude
shorter than the global chain relaxation times.
The ratio $\tau_\text{bind}/\tau_R \approx 0.01$--$0.04$
places the system firmly in the fast-exchange regime
throughout. Compression deepens this condition: $\tau_R$
increases by an order of magnitude across the pressure
range while $\tau_\text{bind}$ remains essentially
unchanged. The contact excess $n_\text{obs}/n_\text{rand}$
encodes the sequence-level charge distribution directly:
P--P depletion from like-charge repulsion, P--H excess
at low pressure from charge complementarity converging
to random mixing above $P \approx 6$~bar, and H--H
contacts reflecting the balance between like-charge
repulsion and G\={o}-mediated globular domain attraction.
The subdiffusive dynamics and broad $P(D)$ distribution
are collective viscoelastic effects, not consequences
of individual long-lived contacts.

\paragraph{Comparison with experiment, model limitations, and outlook.}
The dynamical timescales obtained from the present coarse-grained
model should not be interpreted as quantitatively mapped physical
times. The use of implicit solvent and Langevin dynamics eliminates
explicit solvent-mediated hydrodynamic interactions and modifies the
effective friction experienced by the proteins. In addition, coarse
graining smooths the molecular free-energy landscape and can
accelerate conformational and translational dynamics. Consequently,
a single universal conversion factor between simulation and
experimental time cannot be assumed.

The robust dynamical conclusions of the present study therefore
concern relative trends and the hierarchy of timescales rather than
their absolute mapping to experiment. In particular, intermolecular
contacts relax on a much shorter timescale than whole-chain
conformational relaxation, $\tau_{\rm bind}\ll\tau_R$, while both
translational and conformational dynamics slow systematically with
increasing density. This separation of timescales establishes the
fast-exchange character of the condensate within the present model
without requiring an absolute conversion between simulation and
experimental time.

The neglect of hydrodynamic interactions is also expected to affect
translational diffusion and collective relaxation. Explicit
solvent-mediated hydrodynamic coupling can modify both the magnitude
and chain-length dependence of diffusion and conformational
relaxation, and may therefore change the absolute values of $D$ and
$\tau_R$ reported here. We do not expect this omission to alter the
principal structural conclusions or the qualitative slowing of the
dynamics with increasing packing fraction. Incorporating
hydrodynamic interactions, through either an explicit-solvent
description or an appropriate hydrodynamically coupled
coarse-grained scheme, would be required for a quantitative
comparison of dynamical timescales with experiment.

Finite-size effects provide an additional limitation for the
translational dynamics. The simulations contain 50 ProT$\alpha$
and 40 Histone H1 molecules in periodic cubic boxes whose linear
dimensions vary from approximately 23 to 17 nm across the states
studied. Periodic boundary conditions can modify long-wavelength
collective motion and therefore the absolute translational mobility.
For systems exhibiting normal diffusion, the finite-size dependence
of the translational diffusion coefficient under periodic boundary
conditions can be corrected using the Yeh--Hummer relation
\cite{yeh2004system}. In the present simulations, however, the
center-of-mass motion does not reach a fully Fickian asymptotic
regime; the reported $D$ values are effective mobilities obtained
from finite-time fits to $\mathrm{MSD}(t)=6Dt^\alpha$ with
$\alpha<1$. We therefore do not apply the conventional
Yeh--Hummer correction to these effective diffusion coefficients.
Moreover, the implicit-solvent Langevin description does not
contain the hydrodynamic solvent response on which this correction
is based. We consequently interpret the absolute values of $D$
cautiously, while emphasizing the robust relative result that
mobility decreases systematically with increasing packing fraction
for both protein species.

A second limitation concerns the implicit treatment of ionic
screening. Although electrostatic interactions are screened through
the
Debye--H\"uckel potential, the model does not contain explicit counterions and therefore cannot represent counterion condensation, ion correlations, or counterion-release effects explicitly.

Our earlier work has shown that single-bead hydropathy models
consistently overestimate the radius of gyration for isolated
highly charged IDPs — by approximately 10--20\% for ProT$\alpha$N
($Q_{\rm net} = -43$) and ProT$\alpha$C ($Q_{\rm net} = -40$) —
and that this overestimation is robust across two different
hydropathy scales~\cite{seth2024fine}, consistent with a Manning
condensation picture in which the deviation grows with net
charge~\cite{manning1969limiting}.
In the condensate, however, the near charge-neutrality of the
mixture ($\Delta Q \approx -80\,e$ overall) provides many-body
internal screening absent for an isolated chain, and the
overestimated self-repulsion is partially compensated by the
equally overestimated inter-chain attraction, reducing the
net sensitivity of condensate observables to this approximation.
For the dynamical conclusions, Debye screening at physiological
ionic strength is expected to reduce the P--H attraction and
shorten $\tau_{\rm bind}$ further, increasing the ratio
$\tau_R/\tau_{\rm bind}$ and thereby strengthening the
fast-exchange conclusion.

The general framework may also be useful for other
charge-complementary protein systems with architectures similar to
ProT$\alpha$--H1. One example is the association of positively charged
histones with acidic histone chaperones, where distributed
electrostatic interactions between oppositely charged protein regions
play an important role. For predominantly disordered partners, the
present sequence-resolved HPS description could be applied after
specifying the corresponding sequences and solution conditions,
whereas proteins containing stable folded domains would require an
appropriate structure-based representation analogous to the
native-contact model used here for the globular domain of H1. More
generally, quantitative transfer to a different system would require
validation or reparameterization of the short-range interactions,
electrostatic screening, and any structure-specific terms. Systems
whose association depends strongly on explicit ions, nucleic acids, or
highly specific binding interfaces would require additional degrees of
freedom beyond the present model.

A detailed study of the effects of counter-ions, as well as
the frequency-dependent mechanical response, are important
directions for future work on the ProT$\alpha$--H1 system
and other charge-complementary IDP condensates.

\section{Data Availability Statement}
The data that supports the findings of this study are available from the corresponding author upon reasonable request.
\section{Acknowledgements}
All computations were carried out using the STOKES High-Performance
Computing Cluster at the University of Central Florida.
A.B.\ thanks Robert Best for hospitality at the Laboratory of
Chemical Physics, National Institutes of Health, and gratefully
acknowledges his guidance in setting up the GROMACS implementation
and native contact stabilized G\={o}-type model parameterization of Histone H1 during his sabbatical,
when this work was initiated.
He also thanks Benjamin Schuler for stimulating discussions.
\vfill
% ====================
% References
% ====================
\bibliographystyle{unsrt} 
\bibliography{Master_IDP_References.bib}

@article{uversky2000natively,
  title={Why are "natively unfolded" proteins unstructured under physiologic conditions?},
  author={Uversky, Vladimir N and Gillespie, Joel R and Fink, Anthony L},
  journal={Proteins:Struct., Funct., Bioinf.},
  volume={41},
  number={3},
  pages={415--427},
  year={2000},
  publisher={Wiley Online Library}
}

@article{oldfield2014intrinsically,
  title={Intrinsically disordered proteins and intrinsically disordered protein regions},
  author={Oldfield, Christopher J and Dunker, A Keith},
  journal={Annu. Rev. Biochem.},
  volume={83},
  number={1},
  pages={553--584},
  year={2014},
  publisher={Annual Reviews}
}

@article{olsen2017behaviour,
  title={Behaviour of intrinsically disordered proteins in protein--protein complexes with an emphasis on fuzziness},
  author={Olsen, Johan G and Teilum, Kaare and Kragelund, Birthe B},
  journal={Cellular and Molecular Life Sciences},
  volume={74},
  number={17},
  pages={3175--3183},
  year={2017},
  publisher={Springer}
}

@article{ghosh2022rules,
  title={Rules of physical mathematics govern intrinsically disordered proteins},
  author={Ghosh, Kingshuk and Huihui, Jonathan and Phillips, Michael and Haider, Austin},
  journal={Annu. Rev. Biophys.},
  volume={51},
  number={1},
  pages={355--376},
  year={2022},
  publisher={Annual Reviews}
}

@misc{mobiDB,
  title        = {{MobiDB}: A Database of Intrinsically Disordered Proteins},
  howpublished = {\url{https://mobidb.org/}},
  note         = {Accessed: 2025-06-08}
}

@article{sickmeier2007disprot,
    author = {Sickmeier, Megan and Hamilton, Justin A. and LeGall, Tanguy and Vacic, Vladimir and Cortese, Marc S. and Tantos, Agnes and Szabo, Beata and Tompa, Peter and Chen, Jake and Uversky, Vladimir N. and Obradovic, Zoran and Dunker, A. Keith},
    title = {{DisProt}: the Database of Disordered Proteins},
    journal = {Nucleic Acids Res.},
    volume = {35},
    number = {suppl\_1},
    pages = {D786--D793},
    year = {2006},
    month = {12},
    issn = {0305-1048},
    doi = {10.1093/nar/gkl893}
    
}

@article{dignon2018sequence,
  title={Sequence determinants of protein phase behavior from a coarse-grained model},
  author={Dignon, Gregory L and Zheng, Wenwei and Kim, Young C and Best, Robert B and Mittal, Jeetain},
  journal={PLoS Comput. Biol.},
  volume={14},
  number={1},
  pages={e1005941},
  year={2018},
  publisher={Public Library of Science San Francisco, CA USA}
}

@article{tesei2021accurate,
  title={Accurate model of liquid--liquid phase behavior of intrinsically disordered proteins from optimization of single-chain properties},
  author={Tesei, Giulio and Schulze, Thea K and Crehuet, Ramon and Lindorff-Larsen, Kresten},
  journal={Proc. Natl. Acad. Sci. U. S. A.},
  volume={118},
  number={44},
  pages={e2111696118},
  year={2021},
  publisher={National Academy of Sciences}
}

@article{seth2024fine,
  title={Fine structures of intrinsically disordered proteins},
  author={Seth, Swarnadeep and Stine, Brandon and Bhattacharya, Aniket},
  journal={J. Chem. Phys.},
  volume={160},
  number={1},
  year={2024},
  publisher={AIP Publishing}
}

@article{ashbaugh2008natively,
  title={Natively unfolded protein stability as a coil-to-globule transition in charge/hydropathy space},
  author={Ashbaugh, Henry S and Hatch, Harold W},
  journal={J. Am. Chem. Soc.},
  volume={130},
  number={29},
  pages={9536--9542},
  year={2008},
  publisher={ACS Publications}
}

@book{israelachvili2011intermolecular,
  title={Intermolecular and surface forces},
  author={Israelachvili, Jacob N},
  year={2011},
  publisher={Academic press}
}

@article{akerlof1950dielectric,
  title={The dielectric constant of water at high temperatures and in equilibrium with its vapor},
  author={Akerlof, GC and Oshry, HI},
  journal={Journal of the American Chemical Society},
  volume={72},
  number={7},
  pages={2844--2847},
  year={1950},
  publisher={ACS Publications}
}

@article{borgia2018extreme,
  title={Extreme disorder in an ultrahigh-affinity protein complex},
  author={Borgia, Alessandro and Borgia, Madeleine B and Bugge, Katrine and Kissling, Vera M and Heidarsson, P{\'e}tur O and Fernandes, Catarina B and Sottini, Andrea and Soranno, Andrea and Buholzer, Karin J and Nettels, Daniel and others},
  journal={Nature},
  volume={555},
  number={7694},
  pages={61--66},
  year={2018},
  publisher={Nature Publishing Group UK London}
}

@article{chowdhury2023driving,
  title={Driving forces of the complex formation between highly charged disordered proteins},
  author={Chowdhury, Aritra and Borgia, Alessandro and Ghosh, Souradeep and Sottini, Andrea and Mitra, Soumik and Eapen, Rohan S and Borgia, Madeleine B and Yang, Tianjin and Galvanetto, Nicola and Ivanovi{\'c}, Milo{\v{s}} T and others},
  journal={Proceedings of the National Academy of Sciences},
  volume={120},
  number={41},
  pages={e2304036120},
  year={2023},
  publisher={National Academy of Sciences}
}

@article{muller2010charge,
  title={Charge interactions can dominate the dimensions of intrinsically disordered proteins},
  author={M{\"u}ller-Sp{\"a}th, Sonja and Soranno, Andrea and Hirschfeld, Verena and Hofmann, Hagen and R{\"u}egger, Stefan and Reymond, Luc and Nettels, Daniel and Schuler, Benjamin},
  journal={Proceedings of the National Academy of Sciences},
  volume={107},
  number={33},
  pages={14609--14614},
  year={2010},
  publisher={National Academy of Sciences}
}

@article{FUS_sim,
  author  = {Abyzov, Anton and Blackledge, Martin and Zweckstetter, Markus},
  title   = {Heterogeneous Slowdown of Dynamics in the Condensate of an
             Intrinsically Disordered Protein},
  journal = {J. Phys. Chem. Lett.},
  year    = {2024},
  volume  = {15},
  pages   = {9027--9035},
  doi     = {10.1021/acs.jpclett.4c02142},
  note    = {Reports {$D = 0.26 \times 10^{-3}\,\mathrm{nm}^2/\mathrm{ns}$}
             for FUS-LCD in condensate}
}

@article{Fawzi,
  author    = {Burke, Kathleen A. and Janke, Abigail M. and Rhine, Christy L.
               and Fawzi, Nicolas L.},
  title     = {Residue-by-Residue View of {In Vitro} {FUS} Granules that Bind
               the {C}-terminal Domain of {RNA} Polymerase {II}},
  journal   = {Mol. Cell},
  year      = {2015},
  volume    = {60},
  number    = {2},
  pages     = {231--241},
  doi       = {10.1016/j.molcel.2015.09.006},
  note      = {NMR diffusometry: D = $0.17 \pm 0.02$ um$^2$/s in FUS LC condensate}
}

@article{watanabe2025diffusion,
  title={Diffusion of intrinsically disordered proteins within protein condensates},
  author={Watanabe, Fuga and Akimoto, Takuma and Best, Robert B and Lindorff-Larsen, Kresten and Metzler, Ralf and Yamamoto, Eiji},
  journal={Physical Review Research},
  volume={7},
  number={4},
  pages={043117},
  year={2025},
  publisher={APS}
}

@article{FUS_viscoelastic,
  author  = {Jawerth, Louise and others},
  title   = {Protein condensates as aging Maxwell fluids},
  journal = {Science},
  year    = {2020},
  volume  = {370},
  pages   = {1317--1323},
  doi     = {10.1126/science.aaw4951}
}

@article{Schuler_smFRET,
  author  = {Schuler, Benjamin and Soranno, Andrea and
             Hofmann, Hagen and Nettels, Daniel},
  title   = {Single-molecule {FRET} spectroscopy and
             the polymer physics of unfolded and
             intrinsically disordered proteins},
  journal = {Annu. Rev. Biophys.},
  year    = {2016},
  volume  = {45},
  pages   = {207--231},
  doi     = {10.1146/annurev-biophys-062215-010915}
}

@book{rubinstein2003polymer,
  title={Polymer physics},
  author={Rubinstein, Michael and Colby, Ralph H},
  year={2003},
  publisher={Oxford university press}
}

@article{metzler2014anomalous,
  title={Anomalous diffusion models and their properties: non-stationarity, non-ergodicity, and ageing at the centenary of single particle tracking},
  author={Metzler, Ralf and Jeon, Jae-Hyung and Cherstvy, Andrey G and Barkai, Eli},
  journal={Physical Chemistry Chemical Physics},
  volume={16},
  number={44},
  pages={24128--24164},
  year={2014},
  publisher={Royal Society of Chemistry}
}

@article{humphrey1996vmd,
  title={VMD: visual molecular dynamics},
  author={Humphrey, William and Dalke, Andrew and Schulten, Klaus},
  journal={Journal of molecular graphics},
  volume={14},
  number={1},
  pages={33--38},
  year={1996},
  publisher={Elsevier}
}

@article{go1983theoretical,
  title={Theoretical studies of protein folding},
  author={Go, Nobuhiro},
  journal={Annual review of biophysics and bioengineering},
  volume={12},
  number={1},
  pages={183--210},
  year={1983},
  publisher={Annual Reviews 4139 El Camino Way, PO Box 10139, Palo Alto, CA 94303-0139, USA}
}

@article{manning1969limiting,
  title={Limiting laws and counterion condensation in polyelectrolyte solutions I. Colligative properties},
  author={Manning, Gerald S},
  journal={The journal of chemical Physics},
  volume={51},
  number={3},
  pages={924--933},
  year={1969},
  publisher={American Institute of Physics}
}

@article{Muthukumar2004,
  author  = {Muthukumar, M.},
  title   = {Theory of counter-ion condensation on flexible polyelectrolytes:
             Adsorption mechanism},
  journal = {J. Chem. Phys.},
  volume  = {120},
  pages   = {9343--9350},
  year    = {2004},
  doi     = {10.1063/1.1701839}
}

@article{Muthukumar2017,
  author  = {Muthukumar, M.},
  title   = {50th Anniversary Perspective: A Perspective on
             Polyelectrolyte Solutions},
  journal = {Macromolecules},
  volume  = {50},
  pages   = {9528--9560},
  year    = {2017},
  doi     = {10.1021/acs.macromol.7b01929}
}

@article{yeh2004system,
  title={System-size dependence of diffusion coefficients and viscosities from molecular dynamics simulations with periodic boundary conditions},
  author={Yeh, In-Chul and Hummer, Gerhard},
  journal={The Journal of Physical Chemistry B},
  volume={108},
  number={40},
  pages={15873--15879},
  year={2004},
  publisher={ACS Publications}
}
\end{document}